\documentclass[aps,prb,twocolumn,showpacs,floats,amsmath,amssymb,longbibliography]{revtex4-1}
\usepackage{hyperref}
\usepackage{graphicx,bm}
\usepackage{times}
\usepackage{amsmath,amssymb}
\usepackage{multirow}

\def\moy#1{\left\langle #1 \right\rangle}

\newcommand{\kb}{{\bm k}}

\newcommand{\rb}{{\bm r}}

\begin{document}
\title{Geometry-induced memory effects in isolated quantum systems: Observations and applications}
\author{Chen-Yen Lai and Chih-Chun Chien}
\affiliation{School of Natural Sciences, University of California, Merced, CA 95343, USA.}
\date{\today}

\begin{abstract}
Memory effects can lead to history-dependent behavior of a system, and they are ubiquitous in our daily life and have broad applications.
Here we explore possibilities of generating memory effects in simple isolated quantum systems.
By utilizing geometrical effects from a class of lattices supporting flat-bands consisting of localized states, memory effects could be observed in ultracold atoms in optical lattices.
As the optical lattice continuously transforms from a triangular lattice into a kagome lattice with a flat band, history-dependent density distributions manifest quantum memory effects even in noninteracting systems, including fermionic as well as bosonic systems in the proper ranges of temperatures.
Rapid growth in ultracold technology predicts a bright future for quantum memory-effect systems, and here two prototypical applications of geometry-induced quantum memory effects are proposed:
A cold-atom based accelerometer using an atomic differentiator to record the mechanical change rate of a coupled probe and an atomic quantum memory cell for storing information with write-in and read-out schemes.
\end{abstract}


\maketitle

\section{Introduction}
Memory effects are ubiquitous in our daily life, ranging from the hysteresis loop in magnetization~\cite{JacksonBook,Sun:2003iz} to rechargeable batteries~\cite{LidenBook,Sasaki:2013cd}.
While the former is useful in information technology~\cite{gershenfeld2000physics}, the latter limits the life of portable electronic devices.
When a system exhibits memory effects, its states are history-dependent.
Shape-memory materials~\cite{EbaraBook,CuiSMA} and memory circuit elements~\cite{Pershin:2011ie,Marani15} have been explored and shown potential applications.
Moreover, reproducible microstates in artificial spin-ice driven by a cyclic external magnetic field  have been observed in both experiments and simulations~\cite{Gilbert:2015vx}.
Most applications and understandings of memory effects are based on classical physics, and how memory effects can arise in simple microscopic quantum systems will be the main subject investigated here.

Band theory of electrons \cite{ashcroft1976solid,gershenfeld2000physics} based on a noninteracting picture leads to great success in modern electronics industry.
Here we ask similar questions on whether noninteracting quantum systems can exhibit memory effects and what kinds of applications are out there.
At first look it is counter-intuitive to consider memory effects in noninteracting systems because of a lack of competing time scales and information storage mechanism.
This assertion is supported by previous work on transport in noninteracting quantum systems ~\cite{ChienPS13,Cornean:2013kz} which suggests that no signature of memory effects can be observed in the steady-state current as the system is driven with different time scales.

A mathematical construction based on the idea of a bound state that can jump into and out of a continuum in a quantum-dot system suggests that memory effects may exist in noninteracting quantum  systems~\cite{Cornean:2013kz}, although how such a system can be realized remains unclear.
On the other hand, the tunability of ultracold atom experiments has allowed explorations of quantum effects in both bosonic and fermonic systems in and out of equilibrium~\cite{Bloch:2008gl,Pethick:2010gy,Stoof:2008ho}.
While electrons are always accompanied by Coulomb interactions, interactions of cold-atoms are from two-body scattering, which can be turned off by applying an external magnetic field. Thus,  noninteracting quantum systems are readily available~\cite{Chin:2010kl} and make cold-atom systems particularly suitable for unambiguous demonstrations of memory effects with or without self-interactions.
Indeed, atomic superfluids exhibit hysteresis loops similar to the magnetization~\cite{Eckel:2014gf}, so there is no doubt quantum memory effects can be demonstrated using cold atoms.
The next milestone would be a demonstration of memory effects in noninteracting quantum systems.

Memory effects are closely tied to the dynamics and transport of the underlying systems because different evolution scenarios are needed for revealing the history of the system.
Several transport phenomena in mesoscopic systems have been demonstrated in cold-atom systems~\cite{ChienTranreview}, including quantum ratchet~\cite{Salger:2009ci}, relaxation of fermions in optical lattices~\cite{Schneider:2012ke}, and sloshing motions~\cite{Ott:2004jk}.
Beside analogous phenomena in solid-state systems, cold-atom experiments can further explore spin-imbalanced fermions~\cite{Partridge:2006fq,Zwierlein:2006gb}, inhomogeneous interactions~\cite{Fukuhara:2009bv,Yamazaki:2010en,Blatt:2011dr,Wu:2012jp}, and dynamical transformations of lattice geometry~\cite{Jo:2012bra,Tarruell:2012db}.
The optical kagome lattice realized in Ref.~[\onlinecite{Jo:2012bra}] supports a flat band, which consists of localized states lacking kinetic energy.
Interesting properties of flat-band systems with implications for cold atoms have been intensely studied~\cite{Wu:2007iz,Cooper:2013jg,Paananen:2015ie}, and the presence of a flat band can interfere with quantum transport of other mobile particles~\cite{Chern:2014wg}.

Here, a mechanism of memory effects based on geometrical transformations of a quantum system is analyzed, where the presence of a flat band plays an important role.
The main idea of this work is monitoring the dynamics of a quantum system continuously transformed from one geometry without any flat band into one with at least one flat band.
As demonstrated in Ref.~[\onlinecite{Jo:2012bra}], transforming a triangular lattice into a kagome lattice has been an available technology.
As the time scale of the transformation changes, evidence of quantum memory effects showing different stationary particle distributions will be presented.
To verify the contribution from the flat band, we consider another transformation from a triangular lattice to a square one, where none of the geometries support a flat band, and show that no memory effect can be observed.
In contrast to the bound state mechanism in a quantum dot discussed in Ref.~[\onlinecite{Cornean:2013kz}], the flat-band mechanism finds immediate applications in cold-atom systems and is readily verifiable.
Moreover, our results suggest that the flat-band induced memory effect should be observable in both fermionic and bosonic systems, which will allow for broader applications.

Realizations of memory effects in simple quantum systems will make it possible to use time as an additional control parameter.
The hysteresis of atomic superfluid currents~\cite{Eckel:2014gf} provides an example where the reversal of the current lags behind the stirring.
When compared to typical electronic systems with high Fermi velocity ($\approx\! 10^{6}$m/s) and short tunneling time ($\approx\! 1$ fs)~\cite{ashcroft1976solid}, the relatively slow motion of cold atoms with typical tunneling time scales of $\approx\! 1$ ms \cite{Chien:2012ft} makes them particularly suitable for unveiling mechanisms responsible for quantum memory effects and exploring their applications.
For instance, an accelerometer based on an atomic differentiator, a rate-controlled density valve, and a quantum memory-effect atomic memory (QMEAM) will be proposed here.
Similar ideas can lead to novel quantum devices in atomtronics~\cite{Seaman:2007kx,Pepino:2009jb}, where cold atoms in artificial confining potentials are employed to simulate or complement electronic devices.

This paper is organized as follows.
In Sec.~\ref{sec:kspace}, we model feasible experimental setups for lattice transformations capable of demonstrating quantum memory effects and show a proof-of-principle demonstration that memory effects already exist in noninteracting quantum systems in the thermodynamic limit.
In Sec.~\ref{sec:rspace}, we consider real experimental conditions and simulate lattice transformations in finite-size systems which model experiments more faithfully. Clear demonstrations of geometry-induced memory effects are also observable in finite-size systems.
Sec.~\ref{sec:apps} presents several applications of quantum memory effects and their possible experimental realizations. Sec.~\ref{sec:cons} concludes our study.

\section{Dynamics and memory effects after lattice transformations}\label{sec:kspace}
The Hamiltonian for a two-dimensional noninteracting system in a time-dependent lattice potential is
\begin{equation}\label{eq:H}
\mathcal{H}(t)=-\frac{\hbar^2}{2m}\nabla^2+V(x,y,t)=K+V.
\end{equation}
Following Ref.~[\onlinecite{Jo:2012bra}], we consider $V(x,y,t)=V_0\left[V_{tri}(x,y)+Z\gamma(t,t_r)V_{ramp}(x,y)\right]$, where $V_0$ indicates the lattice depth and $V_{tri}(x,y)$ forms a time-independent triangular lattice with lattice constant $a_L$.
The additional potential energy, $V_{ramp}$, tunes the lattice to a different geometry dynamically.
Here, the ramping time is $t_r$, $Z$ specifies the final relative strength between $V_{tri}$ and $V_{ramp}$, and $0\! =\! \gamma(t\le 0,t_r)\! \le\!  \gamma(t,t_r)\! \le\! \gamma(t\ge t_r,t_r)\! =\! 1$ characterizes the ramp.
The explicit formulations for the triangular lattice and different ramping potentials are summarized in Appendix~\ref{app:FiniteDiff}.

Figure~\ref{fig:contour} shows the triangular lattice with an enlarged unit cell (consisting of site-$A$, $B$, $C$, and $D$) and illustrates a transformation into the kagome lattice with $V_{ramp}=V_{ramp}^{(k)}$ and one into a square lattice with $V_{ramp}=V_{ramp}^{(s)}$.
The most important difference among the triangular, square, and kagome lattices pertaining to this work is that the kagome lattice supports a flat band as summarized in Appendix~\ref{app:band}.
Since the particles in the flat band have no kinetic energy and do not participate in transport, they introduce a different time scale from the tunneling time of mobile particles and make memory effects possible.

\subsection{Band theory analysis}
\begin{figure}[t]
\begin{center}
\includegraphics[width=0.48\textwidth]{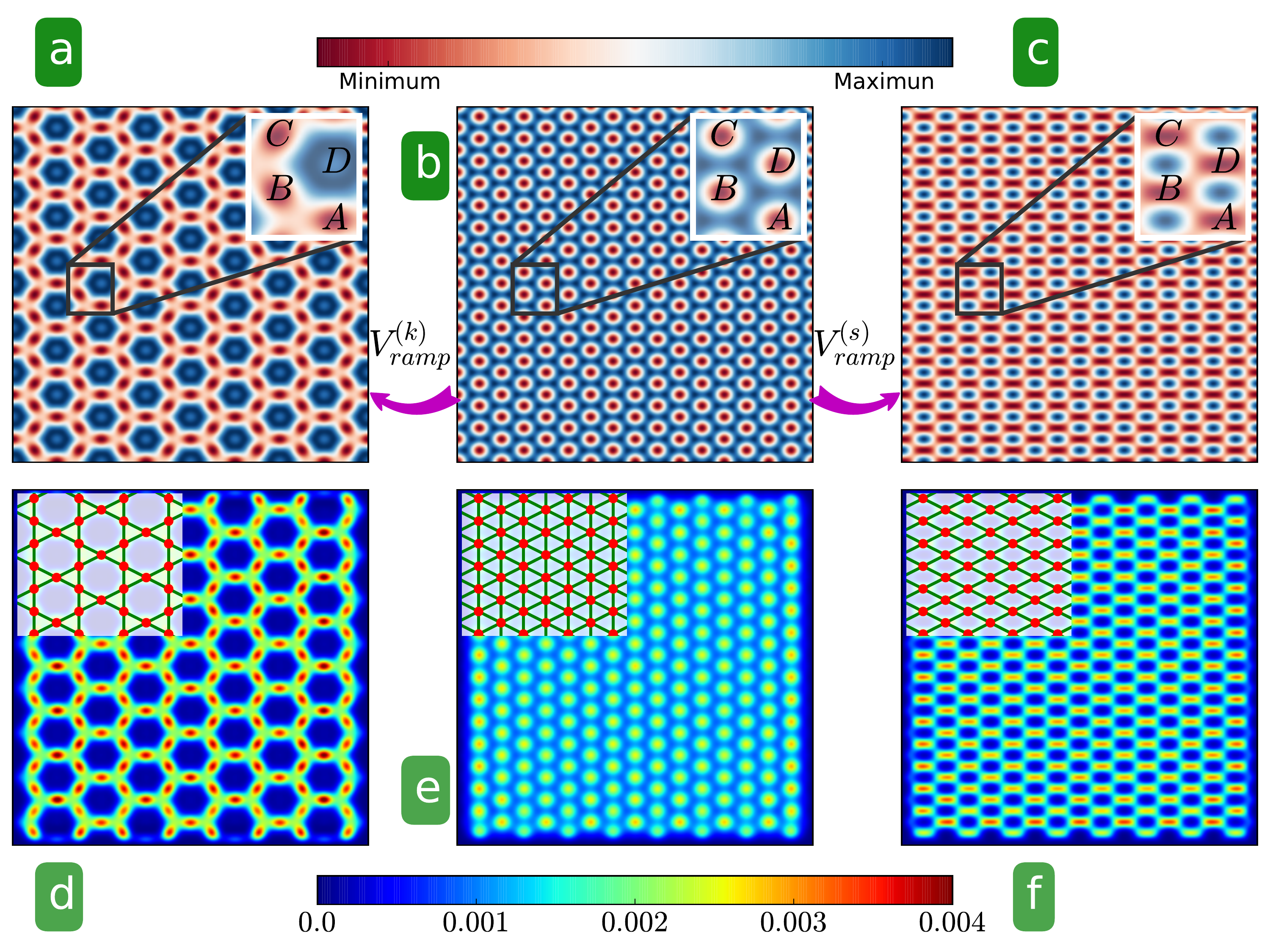}
\caption{(Color online)
Lattice transformations from the triangular lattice (b) to the kagome lattice (a) by tuning the potential on site-$D$, and to the square lattice (c), where a second set of laser increases the barrier between $A$-$D$ and $B$-$C$.
(d)-(f) Snapshots of particle density distributions at different times from simulations:
(e) Initial density distribution on the triangular lattice.
The density distributions at time $t\!=\!t_0/2$ after a geometry quench
to the kagome and square lattices are shown in (d) and (f) respectively.
Details of the lattice structures are shown in the insets.}
\label{fig:contour}
\end{center}
\end{figure}

To clearly demonstrate that the flat band indeed gives rise to memory effects, we first consider the lattice transformation problem in the thermodynamic limit and implement the tight-binding approximation.
The Hamiltonian in the second-quantization form~\cite{DiVentra:2010ks,mahan2000many} is approximated by
\begin{equation}
	\mathcal{H}_{tb}=\sum_{\moy{ij}}\Psi_i^\dagger h_{ij}\Psi_j,
\end{equation}
where
$\Psi_i^\dagger\! =\!
(c_{A,i}^\dagger , c_{B,i}^\dagger , c_{C,i}^\dagger , c_{D,i}^\dagger)$
denotes the creation operator on the sites shown in Fig.~\ref{fig:contour} and $h_{ij}$ only has finite tunneling coefficients $-\bar{t}$ along the links in the lattice.
The lattice Fourier transformation, $\Psi(\kb)\! =\! \sum_{\rb_j}\Psi_je^{i\kb\rb_j}$ with crystal momentum $\kb$, leads to $\mathcal{H}_{tb}\! =\! \sum_{\kb}\Psi^\dagger(\kb)h(\kb)\Psi(\kb)$ with $h(\kb)_{\mu\nu}\! =\! 0$
if $\mu\! =\! \nu$ and $h(\kb)_{\mu\nu}\!=\!-2\bar{t} \cos(\kb\cdot {\bm a}_{\mu\nu})$ if $\mu\!\neq\! \nu$, where $\mu,\nu\! =\! A,B,C,D$ label the sites on an enlarged unit cell.
The vectors ${\bm a}_{\mu\nu}$ link adjacent sites $\mu$ and $\nu$ as shown in Fig.~\ref{fig:contour}.
The unitary transformation, $c_\nu(\kb)\!=\!\sum_\beta (U)_{\nu\beta}c_\beta(\kb)$, diagonalizes the above matrix and leads to  $\mathcal{H}_{tb}(\kb)\!=\!\sum_{\beta,\kb}E_\beta(\kb)c^\dagger_\beta(\kb)c_\beta(\kb)$, where $\beta\!=\!1,2,3,4$ labels the bands in momentum space.

The lattice transformations are then modeled by introducing time-dependent elements of $h(\kb)$ that continuously transform the triangular lattice to either the kagome or square lattice.
Details of the modeling are summarized in Appendix~\ref{app:band}.
The interpolations between those different lattice geometries are not unique, and here we implement a set of smooth transformations which reasonably model realistic situations.
The interpolation we use here will clearly proves the existence of memory effects in noninteracting quantum systems. Other interpolation schemes are expected to only introduce quantitative differences.

The time evolution of the system after the lattice-transformation potential is turned on can be monitored by the single-particle correlation matrix~\cite{Chien:2012ft} $\mathcal{C}_{\mu,\nu}(\kb)\!=\!\moy{ c^\dagger_\mu(\kb)c_\nu(\kb)}$, whose equation of motion (EOM) can be derived in the Heisenberg picture using $i\hbar\partial_{t}(c^\dagger_\mu c_\nu)\! =\! [c^\dagger_\mu c_\nu,\mathcal{H}_{tb}]$. Details of the derivation is shown in Appendix~\ref{app:band}.
For noninteracting systems, the EOM has the same form for fermions and bosons.
The spin-statistics (fermions vs. bosons) is reflected in the initial conditions determined by different distributions, which leads to different time-evolutions.
The EOM for noninteracting systems is solved by the fourth-order Runge-Kutta method~\cite{NRE} with $\delta t\!=\!10^{-2}t_0$, where the time unit in the tight-binding approximation is $t_0\!=\!\hbar/\bar{t}$.
In the following we will set $\hbar\! =\! 1$ and $k_{B}\! =\! 1$.
The population of each band can be evaluated from $\mathcal{C}_{\mu,\nu}(\kb)$ by using
$N_\beta\! =\! \sum_{\kb\in FBZ}\moy{ c^\dagger_\beta(\kb)c_\beta(\kb)}\!=\! \sum_{\kb\in FBZ}\sum_{\mu,\nu} [U(\kb)]_{\beta \mu}\mathcal{C}_{\mu,\nu}(\kb)[U^\dagger(\kb)]_{\nu\beta}$. Throughout the paper, the summation of crystal momentum $\kb$ is over the first Brillouin zone (FBZ) and the summations of $\mu,\nu$ are over $A,B,C,D$ shown in Fig.~\ref{fig:contour}.

\begin{figure}[t]
	\begin{center}
		\includegraphics[width=0.48\textwidth]{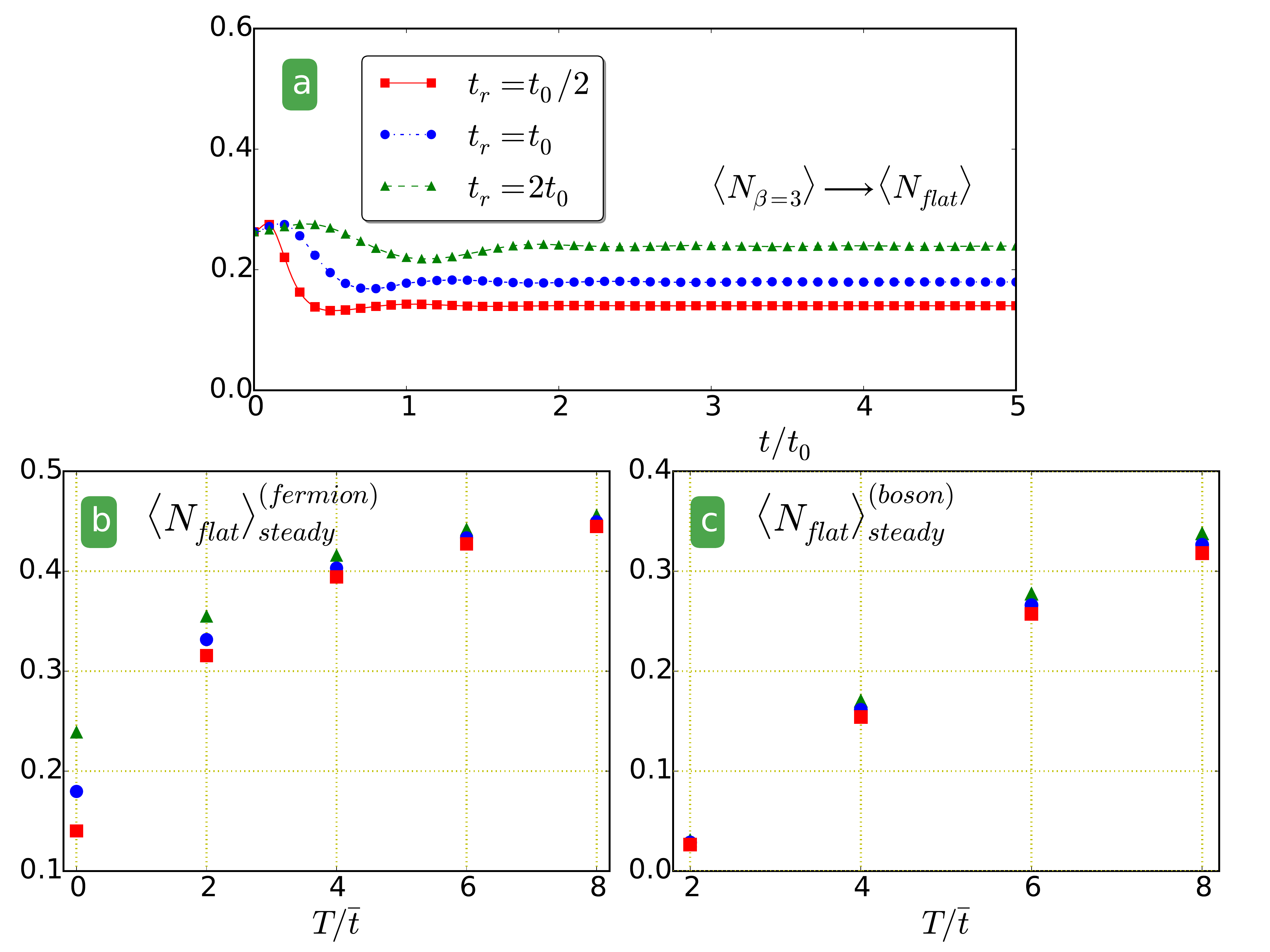}
		\caption{(Color online)
		(a) Averaged fermion density on the third band, $\moy{ N_{\beta=3}}$, after a transformation to the kagome lattice at zero temperature.
		Different symbols stand for different ramping times.
		The initial filling fraction is set to half-filling.
		When the kagome lattice has been reached, $t\! >\! t_r$, the third band becomes a flat band.
		The inset shows the result from a transformation to the square lattice with the same initial condition as the fermion case, and the averaged boson densities reach the same steady-state value despite different transient behavior due to different ramping times.
		(b) Averaged fermion density in a steady state in the kagome flat band when single component fermions undergo lattice transformations at different temperatures and ramping times.
		The three data points on $T/\bar{t}=0$ are determined from the three steady-state values in (a) with the same colors and symbols, respectively.
		(c) The same analysis as (b) for noninteracting bosons.
		}
		\label{fig:KFSK_NiTime}
	\end{center}
\end{figure}

The results for single-component fermions will be presented first.
Here, we consider the system is initially in its ground state at zero temperature and undergoes a transformation from the triangular lattice to the kagome lattice.
In Fig.~\ref{fig:KFSK_NiTime}(a), the averaged particle density in the third band is plotted as a function of time.
Here the initial filling is defined in the triangular lattice, and we choose half-filling to make it easier to observe memory effects.
The dependence of observed memory effects on the initial filling will be explained later.
To clearly contrast memory effects due to different time scales, we consider linear ramping of the lattice potential. Therefore, $\gamma(t,t_r)\! =\! 0$ for $t\! <\! 0$ and  $\gamma(t,t_r)\!=\!\Theta(t\!-\!t_r)\!+\!(t/t_r)\Theta(t_r\!-\!t)$ for $t\ge 0$ with a tunable ramping time $t_r$. Here $\Theta(x)$ is the Heaviside step function.
A transformation to the square lattice follows a similar protocol, and its results are summarized in Appendix.~\ref{app:cp}.

The emergence of a steady state as the population of each band evolves into a stationary value is  crucial to the identification of memory effects. Importantly, steady states are clearly visible in Fig.~\ref{fig:KFSK_NiTime}(a).
The thermal distributions of the steady states are found to deviate from the Fermi-Dirac distribution in thermal equilibrium (shown in Appendix~\ref{app:Thermal}).
The existence of a steady state is nontrivial due to a lack of interactions and dissipation, and the system is usually not expected to equilibrate.
Nevertheless,the system manages to evolve into steady states for both noninteracting fermions and bosons.
Different steady-state values from different ramping times serve as clear evidence for memory effects.
When ramped to a square lattice with different time scales, the final steady states reach the same density distribution on each band despite different transient behavior, as shown in Appendix~\ref{app:cp}.
Therefore, only a density re-distribution takes place but no memory effect can be observed in absence of a flat band. We have checked that no memory effects are observed when the system undergoes a transformation without any flat band at various fillings and finite temperatures.

In stark contrast, Fig.~\ref{fig:KFSK_NiTime}(a) shows that the steady-state populations in the flat band, after the triangular lattice is transformed into the kagome lattice, are different for different ramping times.
This provides an unambiguous proof that geometrical effects, such as a flat band, indeed can induce memory effects in a noninteracting quantum system.
The emergence of a steady state as the initial triangular lattice is transformed into the kagome lattice is highly non-trivial given the fact that, in density-driven transport, the flat band of the kagome lattice can interfere with the emergence of a quasi-steady state~\cite{Chien:2012ft}.
Moreover, the longer the ramping time is chosen, the higher the population in the flat band can be observed.

Since the main contribution to the geometry-induced memory effect is from filling or emptying the flat band after a lattice transformation, an optimal condition would be that the initial population of the would-be flat band is substantially different from the populations of the bands beneath or  above it.
Particles moving into or out of the flat band at different rates after lattice transformations then cause memory effects.
The flat band in the kagome lattice corresponds to the third band in the initial triangular lattice when an enlarged cell containing site-$A$ to $D$ is considered.
Thus, prominent memory effects can be observed if the initial filling of the triangular lattice ranges from $1/2$ to $3/4$ with the details shown in Appendix~\ref{app:FillingBand}.
Similarly, the initial temperature can affect the initial particle distribution and influence memory effects.
As the initial temperature increases, all bands tend to have more uniform distributions and memory effects are washed out.
Finite-temperature effects from different initial conditions are summarized in Fig.~\ref{fig:KFSK_NiTime}(b) showing that memory effect wanes as the initial temperature increases.

\subsection{Non-Interacting Boson}
According to the results of fermions, memory effects due to the flat band is more prominent when the population in the would-be flat band is significantly different from those in nearby bands.
When a Bose-Einstein condensate (BEC) of bosons exists in the ground state, the lowest energy band is macroscopically occupied.
As a consequence, higher-energy bands, including the would-be flat-band, have little influence, so memory effects are not observable.
However, a noninteracting BEC cannot survive at finite temperatures in 2D and the bosons start populating all bands. Thus, the flat-band induced memory effect becomes more observable as temperature increases.
Fig.~\ref{fig:KFSK_NiTime}(c) shows that memory effect starts to emerge in noninteracting bosons at intermediate temperatures.
As temperature gets higher, bosons spread more into higher-energy bands, so memory effects associated with the flat band is more prominent.
However, both bosons and fermions approach the classical distribution at high temperatures.
Therefore, there is no significant population change after lattice transformations above the  quantum-degeneracy temperature, and memory effects vanish in the high-temperature regime.

\section{Finite size system}\label{sec:rspace}
Cold-atom experiments are performed on finite systems. Here we model finite systems more faithfully in real space using a finite difference method~\cite{Vudragovic:2012jz,Bao:2006fg}.
Explicitly, we discretize Eq.~\eqref{eq:H} on a grid with square elements of linear size  $\Delta\!=\!a_L/n_G$, where $n_G$ is the number of grid points.
We present the results with system size $(L_x,L_y)\!=\!(8\sqrt{3}, 16)a_L$ approximately corresponding to a lattice of $N_{lattice}\!=\!16\times16$ sites with $n_G\!=\!20$.
Details of our approach, finite size effect, the choice of $n_G$, and the formulation of ramping potentials are provided in Appendix~\ref{app:FiniteDiff}.

\subsection{Single Component Fermion}
For a finite-size system of noninteracting fermions, we monitor the wavefunction of an initial $T\! =\! 0$ ground state.
Explicitly, the energy spectrum of the triangular lattice Hamiltonian $\mathcal{H}\!=\! K+V_{tri}$ can be obtained numerically. By systematically estimating the number of states $N_f$ according to the filling fraction and system size, we choose the lowest $N_f$ states to form a Fermi sea.
Next we numerically integrate the time-dependent Sch\"{o}dinger equation of each eigenstate, $i\hbar\partial_t\Psi_\nu(x,y,t)\!=\!\mathcal{H}(t)\Psi_\nu(x,y,t)$, by using the fourth-order Runge-Kutta~\cite{NRE} method with a time step $\delta t\!=\! 10^{-3}t_0$.
In the finite system, the unit time is $t_0\!=\!\hbar/E_R$ with the recoil energy $E_R\!=\!\frac{\pi^2\hbar^2}{2ma_L^2}$.
The evolution of the particle density can be monitored by
$n(x,y,t) = \sum_{\nu=1}^N \left|\Psi_\nu(x,y,t)\right|^2.$
In Fig.~\ref{fig:contour}(d)-(f), the particle density contours at different times are shown for the  transformations into the square and kagome lattices with a sudden quench, $\gamma(t, t_r)\!=\!\Theta(t)$.
Initially, particles fill the local minima of the triangular lattice potential in Fig.~\ref{fig:contour}(e).
After a transformation into the kagome or square lattice, particles redistribute accordingly, as shown in Fig.~\ref{fig:contour}(d) and (f).
Since there is no dissipation mechanism in an isolated noninteracting system, particles do not reach thermal equilibrium.

\begin{figure}[t]
\begin{center}
\includegraphics[width=0.48\textwidth]{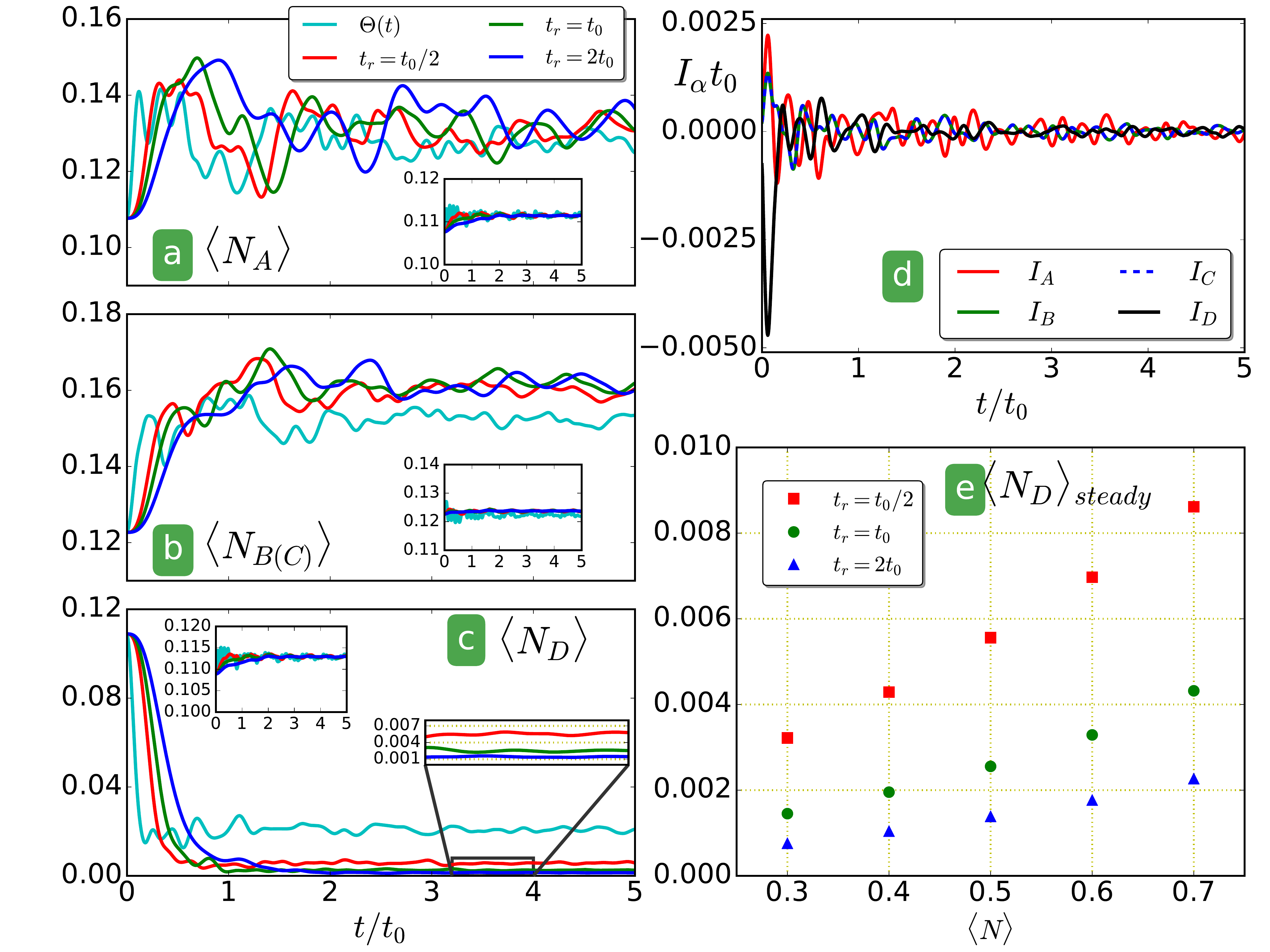}
\caption{(Color online)
(a)-(c) Averaged density on each site versus time for single component fermions experiencing lattice transformations from the triangular to kagome lattices.
The insets of (a)-(c) show the same plots for transformations from the triangular to square lattices.
The quench cases are in cyan color and the linear ramping cases
with different ramping times are $t_r\!=\!t_0/2$ (red), $t_r\!=\!t_0$ (green), and $t_r\!=\!2t_0$ (blue).
(d) Averaged current, $I_\alpha\!=\!\partial_t\moy{N_\alpha(t)}$, versus time for the quench case.
(e) Averaged density, $\moy{N_D}_{steady}$, with different ramping times versus the initial filling fraction, $\moy{N}$.
Slight differences in the initial densities on different sites are due to boundary effects  discussed in Appendix~\ref{app:FiniteDiff}.
}
\label{fig:RFKS_NiTime_T0}
\end{center}
\end{figure}

To better characterize the dynamics of lattice transformations, we define the averaged particle density at site-$\alpha\in\{A, B, C, D\}$ by
\begin{equation}\label{eq:nrs}
\moy{ N_\alpha(t)}=\frac{\sum_{(x, y)\in \alpha}n(x, y, t)}{N_{lattice}}=\frac{N_\alpha}{N_{lattice}}.
\end{equation}
The denominator is the total lattice number and the numerator is the summation of the particle density when its coordinate $(x,y)$ is within a range of the corresponding lattice potential minimum.
The choice of the range does not affect the result qualitatively.
The initial filling fraction is defined by $\langle N\rangle=\sum_{\alpha}\moy{N_\alpha(t=0)}$.
We monitor the particle distributions after transformations into the square and kagome lattices.
Importantly, the non-equilibrium steady states in the thermodynamic limit after lattice transformations manifest themselves here in finite-size systems, as shown in Fig.~\ref{fig:RFKS_NiTime_T0}.
In the long-time limit, the mass current decays to zero, while the particle density on each site approaches its asymptotic value.
Similar to the cases in tight-binding approximation in the thermodynamic limit, the emergence of  quasi steady states in finite systems is crucial because memory effects can be identified by comparing those states after lattice transformations with different time scales.

As shown in Fig.~\ref{fig:RFKS_NiTime_T0}(d), all quasi-steady state currents decay to zero and exhibit no memory effect.
As summarized in Refs.~\onlinecite{Cornean:2013kz,Chien:2013ef}, the current unavoidably misses signatures of memory effects in most cases, and our results align with this observation.
Instead, we investigate the quasi-steady state particle densities and unambiguously identify memory effects associated with lattice transformations where a flat band emerges.
Here the particle density is chosen close to half-filled in the initial triangular lattice.
For a transformation into the square lattice, the averaged particle densities are shown in the insets of Fig.~\ref{fig:RFKS_NiTime_T0}(a)-(c) and exhibit no memory effect.

The averaged particle densities after a transformations into the kagome lattice with different time scales, as shown in Fig.~\ref{fig:RFKS_NiTime_T0}(a)-(c), approach different quasi-steady state values and clearly exhibit memory effects.
In particular, Fig.~\ref{fig:RFKS_NiTime_T0}(c) shows that the residue particle density on site-$D$ highly depends on the ramping time scale.
For linear ramping of the lattice transformation potential, the residual density on site-$D$ varies from less than $0.2\%$ for $t_r\!=\!2t_0$ to nearly $2\%$ for the quench case ($\gamma(t,t_r)\!=\!\Theta(t)$). shown in Fig.~\ref{fig:RFKS_NiTime_T0}(c).
Since the density profiles are readily measurable in cold atoms loaded in optical lattices~\cite{Gemelke:2009ja}, memory effects from the flat-band can be unambiguously observed.
A measurement of the population of the flat-band of the kagome lattice may also be performed by using radio-frequency spectroscopy~\cite{Lee:2007ip,Stewart:2008kt}.

Similar to the band analysis in the thermodynamic limit, flat-band induced memory effects are most significant when the would-be flat band has a different population from its nearby bands.
To quantify the geometry-induced memory effect, we exploit the formation of quasi-steady states after lattice transformations.
Explicitly, we define and evaluate the post transient time-average of the particle density on site-$D$,
\begin{equation}
	\moy{ N_D}_{steady} = \frac{\sum^{T_f}_{t=T_i}\delta t\moy{ N_D(t)}}{T_f-T_i}.
\end{equation}
Here we chose $T_f\! =\! 5t_0$ and $T_i\! =\! 3t_0$, and there is no qualitative  difference from other choices since the system has reached a quasi-steady state.
For a selected initial filling, we interpret the difference between residual particles on site-$D$ in the quasi-steady states from different ramping times as the strength of memory effect and show the result in Fig.~\ref{fig:RFKS_NiTime_T0}(e) for a linear ramping function with different ramping time $t_r$.
The difference in $\moy{N_D}_{steady}$ due to different ramping times increases as the filling fraction increases, which indicates stronger memory effect.

To summarize, atoms in the flat band lack kinetic energy and do not contribute to transport directly, but filling and emptying atoms in the flat band introduce a different time scale other than the hopping time of mobile atoms.
The observed memory effects are from a competition between different time scales, and the ability to retain atoms in the flat band after a lattice transformation because of the emergence of non-trivial steady states allows identification of memory effects.

\subsection{Condensate dynamics of weakly-interacting bosons}
In contrast to fermions, single-species bosons can self interact.
The Gross-Pitaevskii (GP) equation provides a suitable description of the condensate of weakly-interacting bosons far below its BEC transition temperature, and its generalization to time-dependent systems are straightforward~\cite{Vudragovic:2012jz,Pethick:2010gy,Stoof:2008ho}.
In the GP equation, the condensate is described by a wave function $\Phi(x,y,t)$ and
\begin{equation}
\left[-\frac{\hbar^2}{2m}\nabla^2+V(x,y,t)+gN_b|\Phi|^2\right]\Phi = i\hbar\partial_t\Phi ,
\end{equation}
where $N_b$ is the number of bosons. Here we follow Ref.~[\onlinecite{Cerimele:2000hs}] to normalize the wavefunction, $\int dxdy |\Phi(x,y)|^2\!=\!1$.
The coupling constant $g\!=\! 4\pi\hbar^2a_s/m$ is determined by the two-body $s$-wave scattering length $a_s$.
In the following we implement the finite-difference method to investigate condensate dynamics and set $a_s\!=\! (3/\pi)a_L$.

\begin{figure}[t]
\begin{center}
\includegraphics[width=0.48\textwidth]{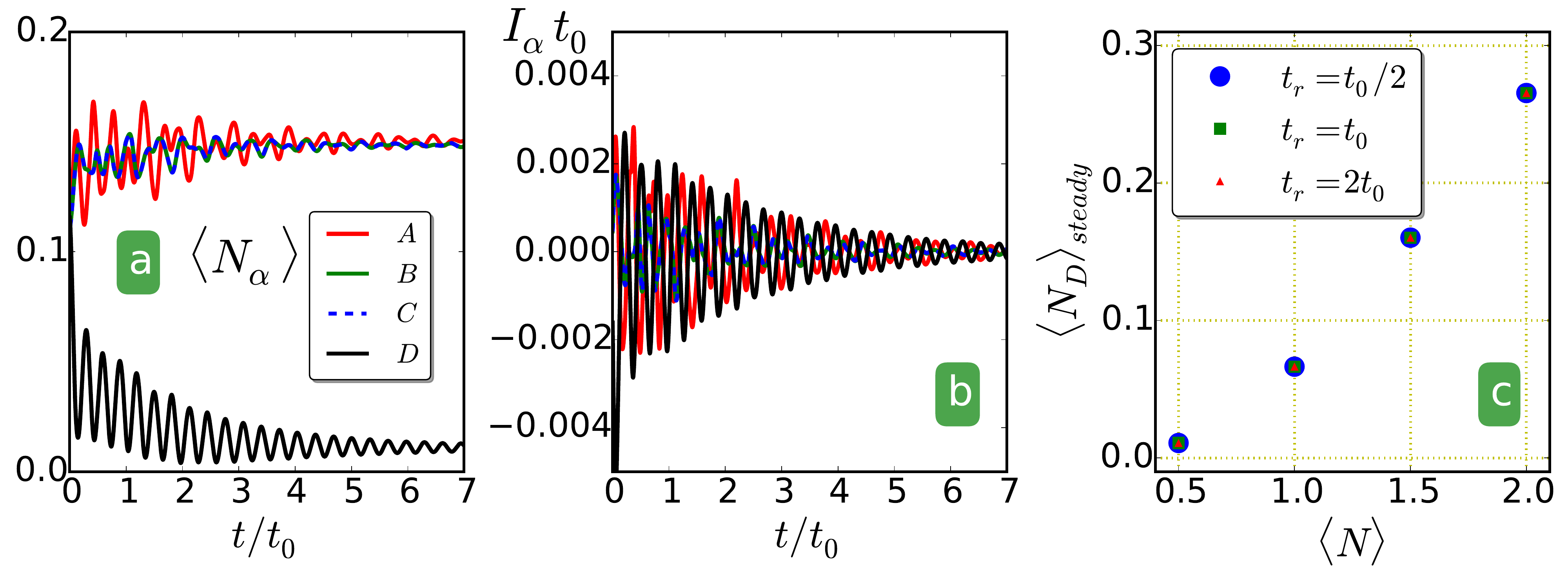}
\caption{(Color online)
(a) Averaged density of noninteracting bosons versus time in a quench from the triangular to kagome lattices.
(b) Averaged current, $I_\alpha\!=\!\partial_t\moy{ N_\alpha(t)}$, of noninteracting bosons versus time.
(c) The same plot as Fig.~\ref{fig:RFKS_NiTime_T0}(e) for weakly-interacting bosons.
Here the points collapse with each other, so no observable memory effect is present.
}
\label{fig:RBK_NDFilling_quench}
\end{center}
\end{figure}

The particle density is given by $n(x,y,t) \!=\! N_b\left|\Phi(x,y,t)\right|^2$.
In Fig.~\ref{fig:RBK_NDFilling_quench}(a) and (b), we show the averaged density and current in each site after a quench into the kagome lattice.
When compared to the fermion case, it takes a longer time for weakly-interacting bosons to reach a quasi-steady state.
The emergence of a quasi steady state in interacting bosons has been studied in Ref.~[\onlinecite{Peotta:2014woa}], and here quasi-steady states after different lattice transformation protocols allow us to detect memory effects.

In the presence of Bose-Einstein condensation, the lowest energy state is macroscopically occupied while other states have relatively small weight.
Since there are two dispersive bands beneath the flat band in the kagome lattice, during and after the lattice transformation, the condensate is less likely to spread into higher bands even when the filling fraction is large.
Observation of flat-band induced memory effects is thus unlikely in weakly interacting bosons in its condensed phase.
Fig.~\ref{fig:RBK_NDFilling_quench}(c) shows the time average of particle density on site-$D$.
One can see that the long-time averaged densities are the same regardless of different ramping times, and there is no observable change as the initial filling increases.
Stronger interaction strength has also been tested and there is no observable flat-band induced memory effect in the condensate.

We remark that memory effects due to the phase stiffness of superfluids have been observed in the  hysteresis loops of bosonic atoms in a ring-shape geometry~\cite{Henderson:2009eo}.
The flat-band induced memory effects discussed here, although not observable in the superfluid phase due to BEC in the ground state, apply to both noninteracting bosons and fermions and complement the memory effects due to long-range orders in superfluids.

\section{Applications}\label{sec:apps}
On the experiment side, the triangular, square, and kagome optical lattices have all been realized~\cite{Jo:2012bra,Tarruell:2012db} and loaded with bosonic atoms.
Feasibility of dynamical ramping is also presented in Ref.~[\onlinecite{Jo:2012bra}].
While measurements of the population in a selected band can be challenging in solid-state materials, real space density distributions in cold-atoms can be mapped out by in-situ imaging~\cite{Gemelke:2009ja}, novel microscopic technique based on scanning electron microscopy~\cite{Gericke:2008jw}, and nondestructive fluorescence imaging~\cite{Patil:2014jv}.
Moreover, the population of a flat band can be measured by time-of-flight experiments~\cite{Wang:2014ke}. A recently developed technique called compensated optical lattice~\cite{Hart:2015ex} can cool fermionic atoms to nearly $7\%$ of the Fermi temperature~\cite{Duarte:2015cn} and bring the system virtually to its ground state.
Combining those imaging and cooling techniques with the possibilities of tuning the lattice geometry, here we propose two applications of geometry-induced memory effects in atomic devices.

\begin{figure}[t]
\begin{center}
\includegraphics[width=0.50\textwidth]{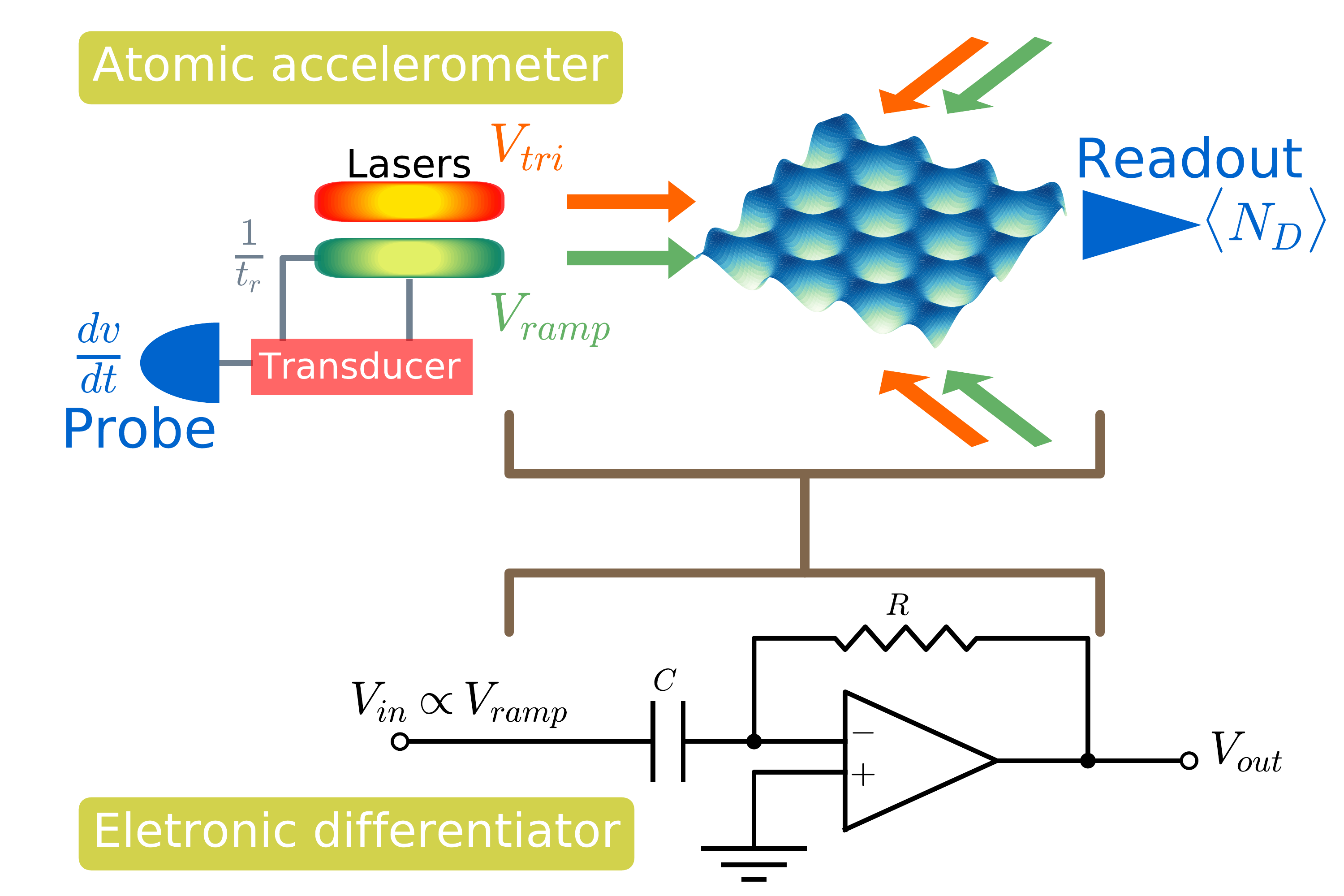}
\caption{(Color online)
Illustration of a memory-effect accelerometer.
The probe receives velocity changes as its signal.
A transducer ramps up the laser for transforming the lattice geometry according to the acceleration received by the probe.
The readout reveals different densities on $D$-site, $\langle N_D\rangle$, due to different accelerations.
The atomtronic part is an analogue of the conventional electronic differentiator shown in the lower panel, where $\langle N_D\rangle$ plays the role of the output signal $V_{out}=-RC(dV_{in}/dt)$.
}
\label{fig:accel}
\end{center}
\end{figure}

The first application is a prototypical atomic differentiator, whose comparison with a conventional electronic differentiator is shown in Fig.~\ref{fig:accel}.
A basic electronic differentiator needs three elements: an operational amplifier, a capacitor, and a resistor \cite{gershenfeld2000physics}.
A faithful analogue of a capacitor using cold atoms can be challenging because atoms are charge neutral and interact via scattering, and an atomic operational amplifier is an even greater challenge.
Instead of building the atomic analogues of all three elements of an electronic differentiator and operating them in a similar fashion, the atomic differentiator based on memory effects can directly reflect the rate of input voltage, which determines the rate of transformation of lattice geometry, by measuring the remaining $D$-site density in the kagome lattice.

Utilizing the atomic differentiator, a cold-atom based accelerometer is also proposed and shown in Fig.~\ref{fig:accel}, where a probe is connected to a piezoelectric piece that controls the laser responsible for the lattice transforming potential $V_{ramp}$.
When the probe is accelerated, a change of voltage from the piezoelectric piece ramps up $V_{ramp}$ and transforms an initially loaded triangular lattice into a kagome lattice.
The remaining density on $D$-sites is sensitive to the rate at which $V_{ramp}$ was switched on, hence it records the acceleration in the probe.
One may, alternatively, record the signals from the transducer and use a conventional electronic differentiator or a computer to process the signals and extract the rate of change.
In contrast to electronic systems using multiple devices to achieve the goal, the proposed atomic device can automatically record, in the remaining $D$-site density, the rate of change sensed by the probe.

Since memory effects in the kagome lattice are most prominent if the ramping time is about the same order of magnitude as the tunneling time, this type of accelerometers will be more efficient in probing changes in the ms-scale.
We remark on a difference between an accelerometer using electronic devices and the cold-atom based accelerometer.
While the electronic accelerometer can operate continuously in time, the cold-atom based accelerometer needs to be refreshed constantly by resetting the lattice back to the triangular lattice and bringing the cold atoms to the ground state, which can be achieved by recent advance in preparing and manipulating small-scale cold-atom systems \cite{Lester:2015if,Beugnon:2007eg,Weiss:2004fk,Beugnon:2007eg}.
Nevertheless, an electronic differentiator loses its output when the input voltage is switched off, but the atomic differentiator or accelerometer keeps its latest output in the remaining density indefinitely.

A second application is to actively control the remaining density using the rate of change of lattice transformation.
We first propose a rate-controlled atomic memvalve, where the ramping potential $V_{ramp}$ controlled by laser intensity is switched on at different rates to tune the remaining density $N_D$ on the $D$-sites of the kagome lattice.
Different ramping rates then lead to different $N_D$, which can be read out using the above-mentioned imaging techniques.
After one operation, $V_{ramp}$ can be turned off and the triangular lattice may be  replenished with fresh cold-atoms for another cycle.

The mechanism behind the memvalve is remotely analogous to an electronic transistor \cite{HorowitzBook}, where a gate voltage controls the current from the source to the drain.
In the memvalve, however, the control is the rate of ramping up $V_{ramp}$. A comparison of the memvalve and transistor is summarized in Table~\ref{tab:memvalve}.
Recent experimental progress in rapid loading and manipulating cold atoms~\cite{Lester:2015if} and portable cold-atom technology~\cite{Rushton14} could make the accelerometer and rate-controlled memvalve realistic in the near future.
\begin{table}[h]
\begin{tabular}{|c|c|c|c|c|c|c|c|}
\hline
\multicolumn{4}{|c|}{
	\begin{minipage}{.2\textwidth}
		\includegraphics[width=\linewidth]{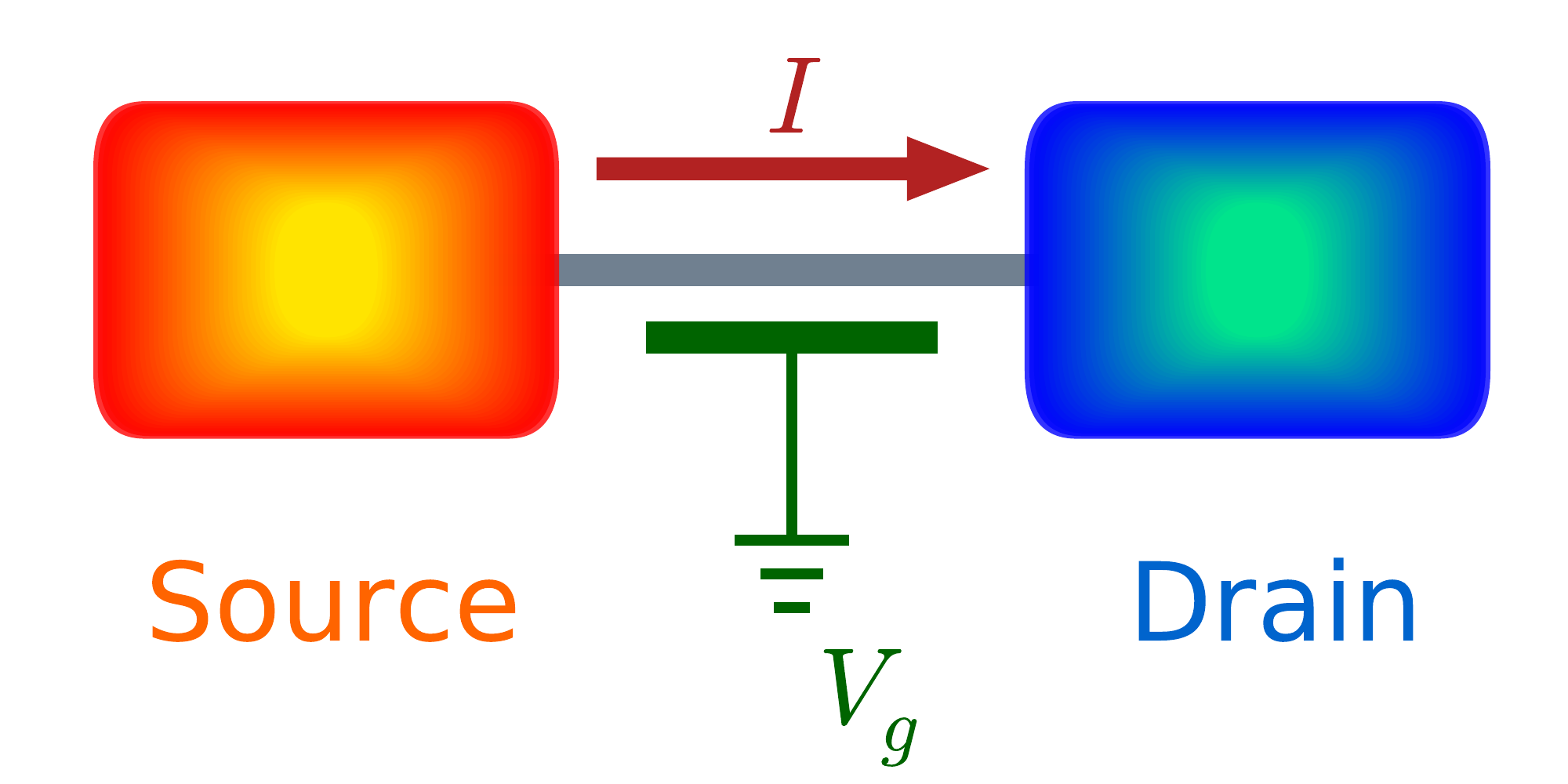}
	\end{minipage}
}                              &
\multicolumn{4}{c|}{
	\begin{minipage}{.2\textwidth}
		\includegraphics[width=\linewidth]{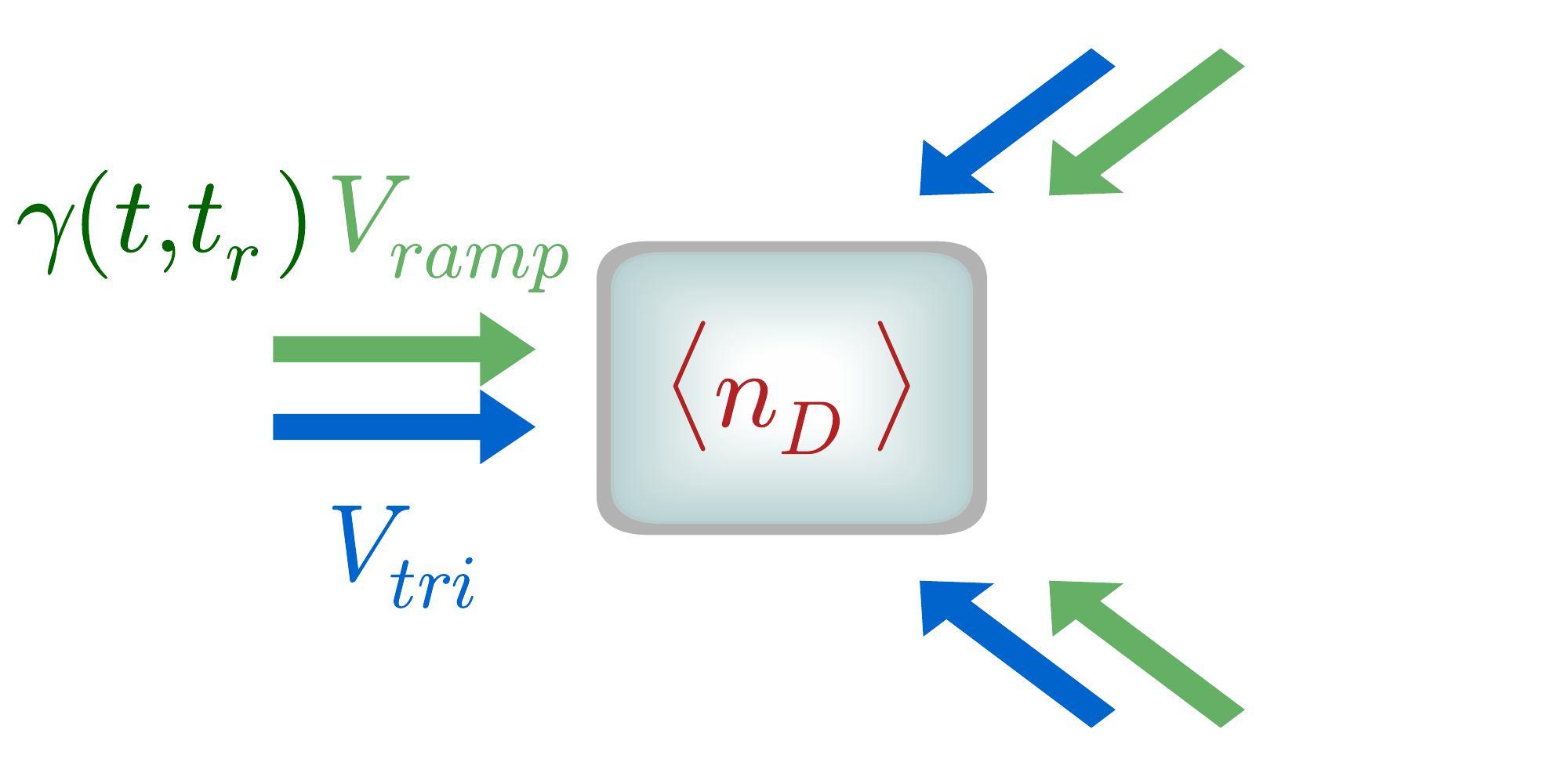}
	\end{minipage}
}                          \\ \hline
\multicolumn{4}{|c|}{Transistor}                              & \multicolumn{4}{c|}{Rate-controlled Memvalve}                          \\ \hline
\multicolumn{2}{|c|}{Control}  & \multicolumn{2}{c|}{Output}  & \multicolumn{2}{c|}{Control} & \multicolumn{2}{c|}{Output}  \\ \hline
\multirow{2}{*}{Voltage} & $V_1$ & \multirow{2}{*}{Current} & $I_1\!>\!I_c$ &  \multirow{2}{*}{Rate}  & $1/t_{r1}$ & \multirow{2}{*}{Density} & $N_{D1}\!>\! N_c$ \\ \cline{2-2} \cline{4-4} \cline{6-6} \cline{8-8}
                         & $V_2$ &                          & $I_2\!<\!I_c$ &                        & $1/t_{r2}$ &                          & $N_{D2}\!<\! N_c$ \\ \hline
\multicolumn{2}{|c|}{
	\begin{minipage}{.1\textwidth}
		\includegraphics[width=\linewidth]{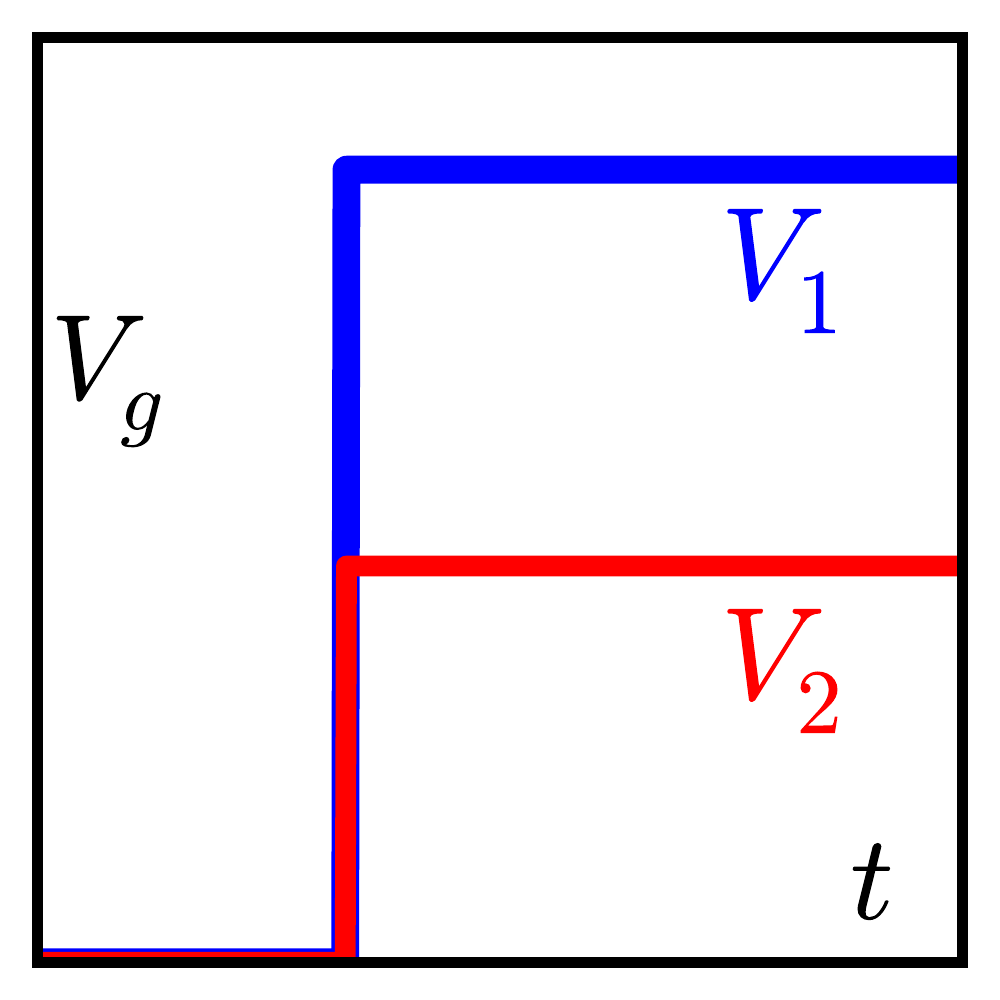}
	\end{minipage}
}
&
\multicolumn{2}{c|}{
	\begin{minipage}{.1\textwidth}
		\includegraphics[width=\linewidth]{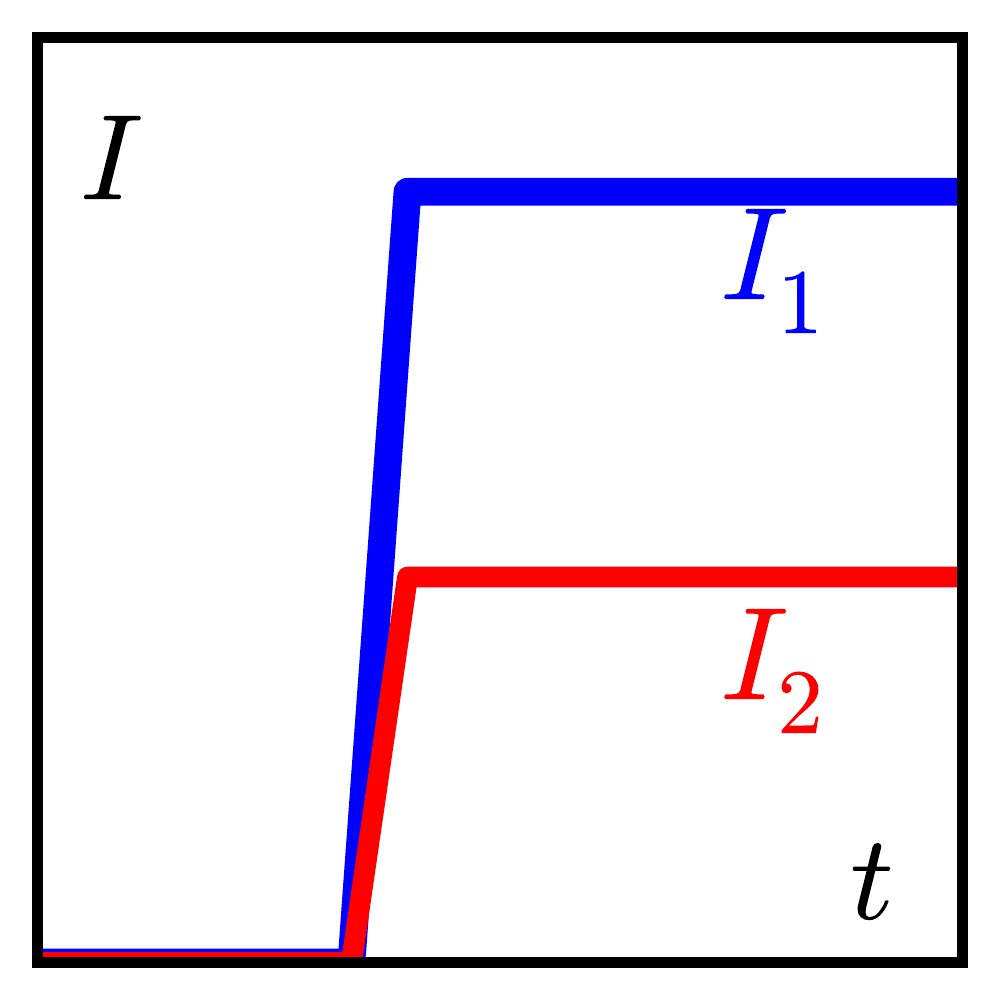}
	\end{minipage}
}
&
\multicolumn{2}{c|}{
	\begin{minipage}{.1\textwidth}
		\includegraphics[width=\linewidth]{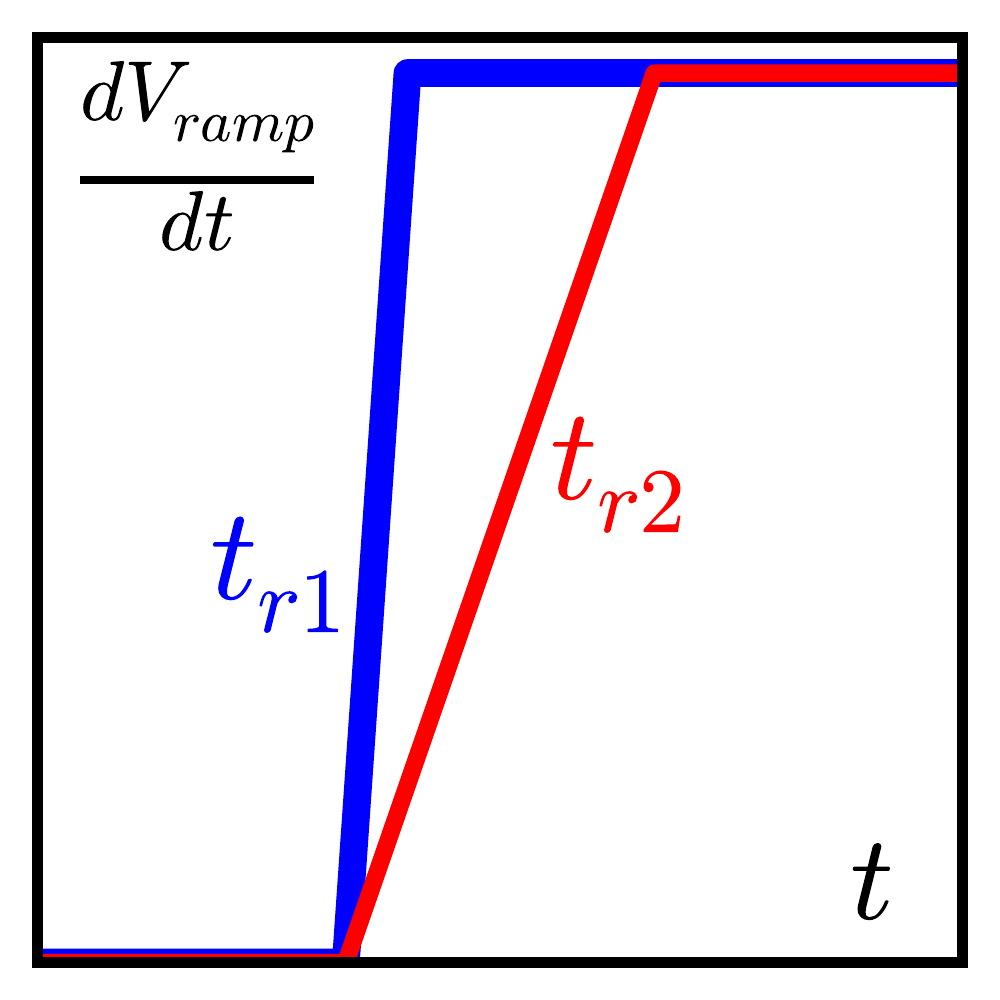}
	\end{minipage}
}
& \multicolumn{2}{c|}{
	\begin{minipage}{.1\textwidth}
		\includegraphics[width=\linewidth]{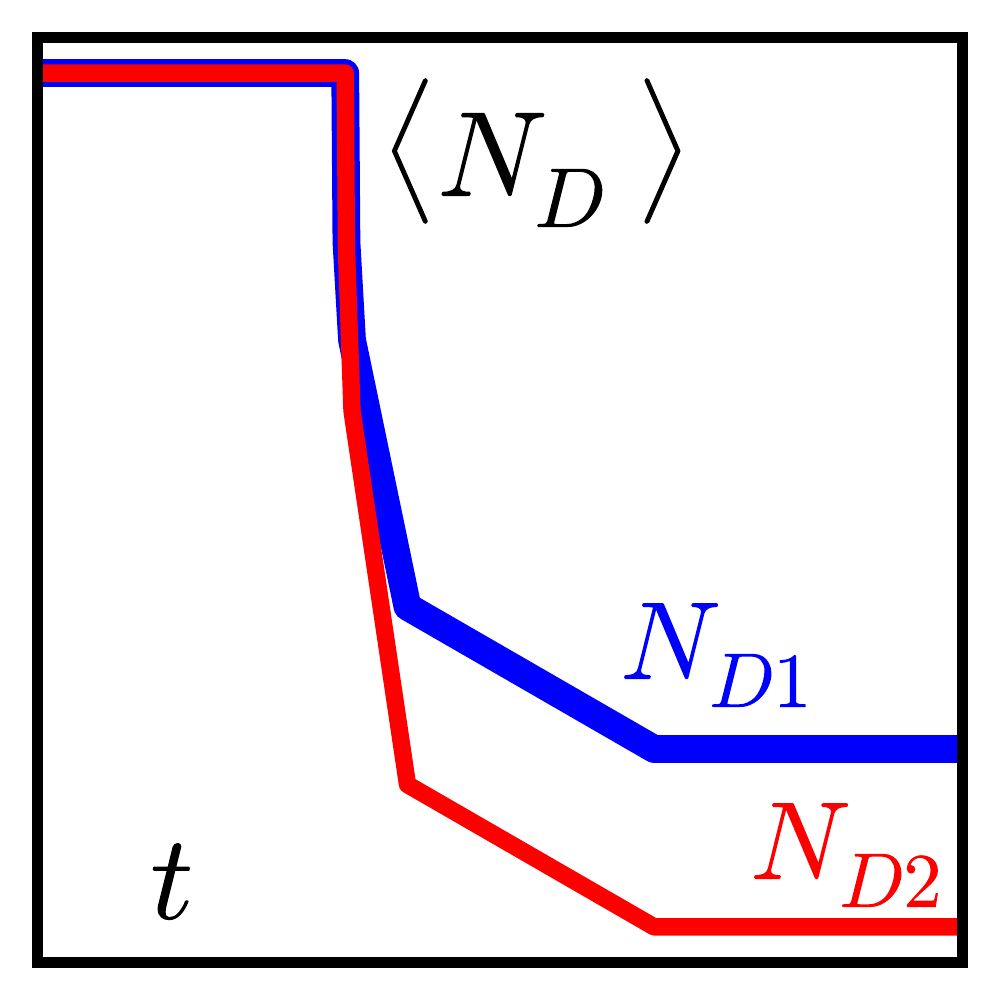}
	\end{minipage}
} \\ \hline
\end{tabular}
\caption{(Color online) Comparison between an electronic transistor and an atomic memvalve.
Left column: An electronic transistor uses the gate voltage $V_g$ to control the output current $I$.
Two sets of input voltages $V_{1,2}$ and the corresponding output currents $I_{1,2}$ represent a binary signal.
Right column: The rate-controlled memvalve takes different lattice-transforming rates, $1/t_{r}$, to control the laser which transforms the lattice geometry to a kagome lattice.
Two different rates lead to two different steady-state $D$-site densities, which also represent a binary signal.
}
\label{tab:memvalve}
\end{table}

\begin{figure}[t]
  \begin{center}
  \includegraphics[width=0.49\textwidth]{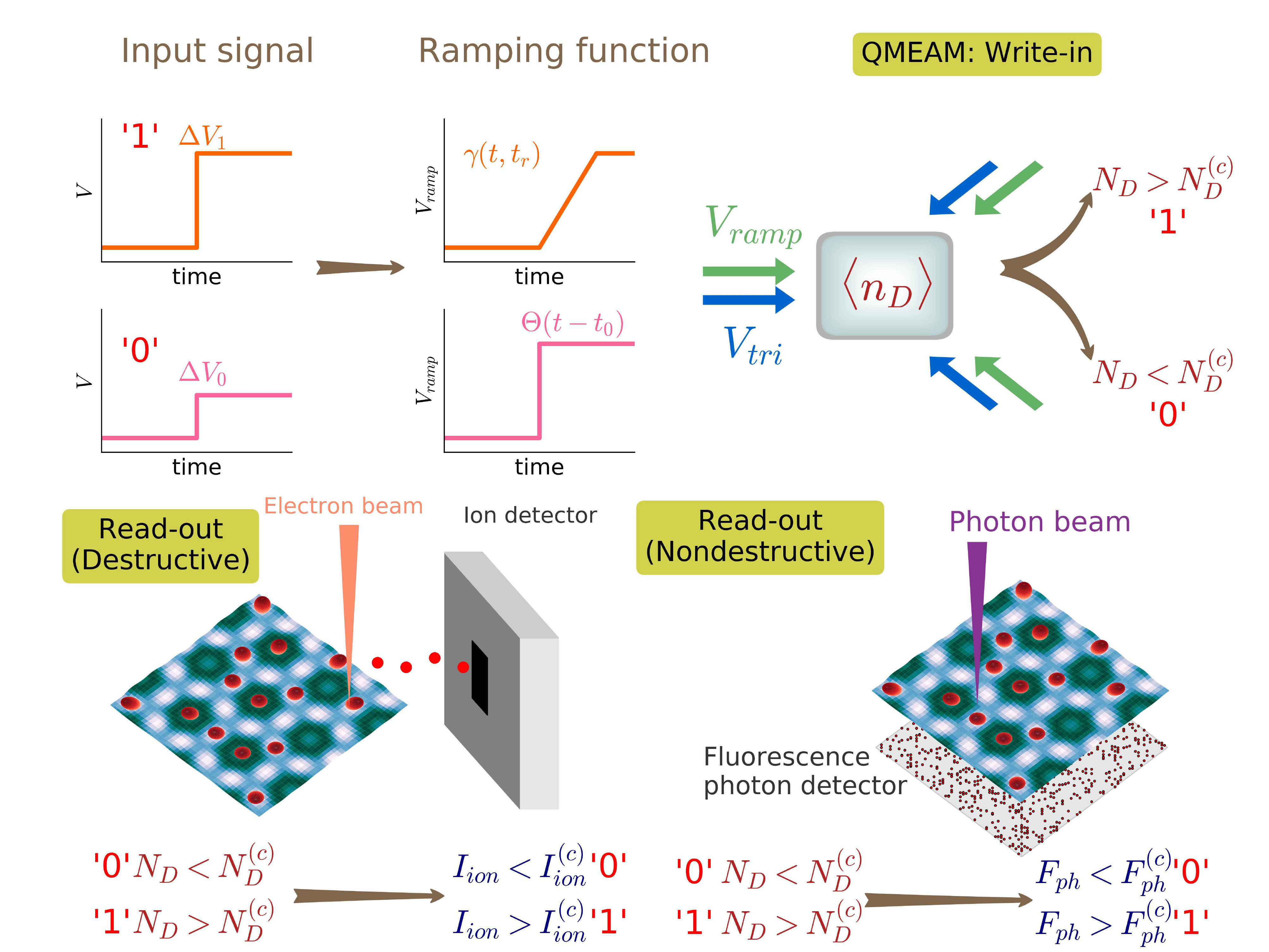}
  \caption{(Color online)
	Quantum memory effect atomic memory (QMEAM). One bit of information can be stored in a memory  cell after a lattice transformation.
	The write-in scheme converts an incoming voltage signal into different rates of ramping the lattice potential $V_{ramp}$.
	The information is then saved as a high or low density ($\langle N_D\rangle$) state.
	Two read-out schemes can be used:
	(lower left) A destructive electron beam converts atoms to an ion current and an ion detector measures the ion current $I_{ion}$ as the output signal.
	(lower right) A nondestructive photon beam induces fluorescence by bringing the atoms out of the dark state and a photo-detector measures the intensity $F_{ph}$ of the fluorescence as the output signal.
	The superscript $c$ stands for a calibrated value to distinguish the two states of one bit. Both $I_{ion}$ and $F_{ph}$ can be amplified and converted to voltage signals.
	}
  \label{fig:qmeam}
  \end{center}
\end{figure}

Based on the memvalve, we propose a quantum memory effects atomic memory (QMEAM) summarized in Fig.~\ref{fig:qmeam}.
Due to the emergence of the steady states after lattice transformations, the remaining $D$-site density does not change in later times.
Thus, this memory element in principle is non-volatile.
The information denoted by "1" and "0" are stored as higher and lower $D$-site densities, respectively.
A proposed write-in process to store information into a QMEAM and two proposed read-out schemes, one destructive and one non-destructive, to retrieve information out of a QMEAM are illustrated in  Fig.~\ref{fig:qmeam}.
Given an input of one bit of information "1" (or "0") represented by a high (or low) voltage, we first send the signal to a single slope analog-to-digital converter architecture~\cite{gershenfeld2000physics,kester2005data}, where the two values of the input voltage translate into two different rates to ramp the laser that transforms the optical lattice from a triangular one to a kagome one.
Since a faster transformation leads to higher remaining $D$-site density, the information is stored as high (or low) $\langle N_{D}\rangle$ in the QMEAM.

Invasive schemes for storing information in lattice atomic systems, such as using the input voltage to control an electron beam to remove atoms from an optical lattice \cite{Gericke:2008jw} and storing information as the overall density difference, are possible.
In contrast, the QMEAM is a number-conserving scheme since the atoms only rearrange themselves after a lattice transformation, and when combined with a non-destructive read-out scheme discussed later, the QMEAM does not need to refill its atoms after one cycle of operation.
Since accidental atomic losses in optical lattices can be made small, the memory cell is non-volatile and the duration of storage is limited by the holding time of atoms in the potential~\cite{Bloch:2008gl,Moore:2015kd}.
Moreover, in Sec.~\ref{sec:rspace} we showed that quantum memory effects are observable for a lattice system with $N_{lattice}\!=\!16\times16$ sites.
This implies that the scalability of the QMEAM is promising.

The read-out schemes for retrieving the information in a memory cell are shown in the lower part of Fig.~\ref{fig:qmeam}.
One destructive scheme is to use an electron beam as a scanning microscope~\cite{Gericke:2008jw} to pump the particles out of the system and convert them to an ion current proportional to the particle density where the electron beam is focused on.
The resulting ion current from a selected $D$-site determines the information by comparing the magnitude of the ion current to a calibrated value.
At the end of the operation, the atoms are removed from the optical lattice and the originally stored information is destroyed.

To keep the originally stored information intact after a read-out operation, we propose a second read-out scheme utilizing all-optical measurements as shown in Fig.~\ref{fig:qmeam}.
In Ref.~[\onlinecite{Patil:2014jv}], a pair of counter propagating photon beams are employed. One pumps the atoms into a dark state and the other fluorescence beam brings the atoms out of the dark state.
The resulting Raman fluorescence image not only reveals the real space density, but also preserves the spatial locations of the atoms and leave them in the vibrational ground state~\cite{Patil:2014jv}.
The light-induced particle loss in the fluorescence measurement is controllable and can be minimized.
For the two read-out schemes, the ion current or the fluorescence intensity can be amplified and converted to voltage signals if desired.

\section{Conclusions}\label{sec:cons}
Memory effects are shown to exist in noninteracting quantum systems when a flat band emerges.
Different density distributions following different ramping times serve as direct evidence for memory effects, which are readily observable in fermionic systems at low temperatures and bosonic systems at intermediate temperatures.
Since flat-bands can be found in a variety of lattice geometries, including saw-tooth~\cite{Zhang:2015vg}, cross stitch~\cite{Flach:2014cm, Metcalf:2015wj}, photonic rhombic lattice~\cite{Mukherjee:2015uf}, and the recently realized optical or photonic Lieb lattice~\cite{Taie15,Vicencio:2015ks,Mukherjee:2015iu}, geometry-induced memory effects may also be studied in those systems.
On the other hand, photon-induced memory effects have been claimed in solid-state materials exhibiting structural transitions~\cite{Mikailov:2006gg,Panich:2008cs}, and the tunability of optical-lattice geometry can provide opportunities for bridging our understanding of memory effects in material science and ultracold atoms.

The geometry-induced quantum memory effect is less prominent in the presence of BEC, but phase-stiffness induced memory effects exhibiting hysteresis of the vorticity have been observed in interacting bosons in the superfluid phase~\cite{Eckel:2014gf}.
Given the variety of memory effects and the broad range of their applications, including the accelerometer, memvalve, and QMEAM discussed here, novel quantum devices utilizing memory effects are promising in the emerging field of atomtronics~\cite{Pepino:2009jb,Seaman:2007kx}.

\acknowledgments
We thank Gia-Wei Chern, Massimiliano Di Ventra, Dan Stamper-Kurn, Michael Zwolak, and YangQuan Chen for useful discussions.

\appendix

\section{Band Theory and Dynamics}
\subsection{Model and Method}\label{app:band}
In the thermodynamic limit, the system is approximated by the one-band tight binding model
\begin{equation}
	\mathcal{H}_{tb}=\sum_{\langle ij\rangle}\Psi_i^\dagger h_{ij}\Psi_j,
\end{equation}
 where
$
\Psi_i^\dagger\!=\!
( \begin{array}{cccc}
c_{A,i}^\dagger & c_{B,i}^\dagger & c_{C,i}^\dagger & c_{D,i}^\dagger
\end{array})
$ is the creation operators on each site.
From Fig.~\ref{fig:contour}, the $A$-$B$ link is connected by a vector ${\bm a}_{AB}=(-\sqrt{3}/2, 1/2)a_L$, and the rest of the links can be determined as well.
After performing the Fourier transformation $\Psi(\kb)=\sum_{\rb_j}\Psi_je^{i\kb\rb_j}$, we are able to obtain the Hamiltonian in momentum space as  $\mathcal{H}_{tb}=\sum_{\kb}\Psi^\dagger(\kb)h(\kb)\Psi(\kb)$, where the elements of $h(\kb)$ are $h(\kb)_{\eta\xi}=0$ if $\xi=\eta$ and $h(\kb)_{\eta\xi}\!=\!-2\bar{t} \cos({\kb\cdot {\bm a}_{\eta\xi}})$ if $\eta\neq\xi$ and $\eta,\xi=A,B,C,D$.
During the dynamical ramping to the kagome or square lattice, $h^{(k/s)}(\kb)$ becomes time dependent. Explicitly,
\begin{equation}
	h^{(k)}(\kb)=\left(\begin{array}{cccc}
	0 & h_{AB} & h_{AC} & h^{(k)}_{AD} \\
	h_{AB} & 0 & h_{BC} & h^{(k)}_{BD} \\
	h_{AC} & h_{BC} & 0 & h^{(k)}_{CD} \\
	h^{(k)}_{AD} & h^{(k)}_{BD} & h^{(k)}_{CD} & h^{(k)}_{DD}
	\end{array}\right) ,
\end{equation}
 where $h^{(k)}_{\eta D}(t)\!=\!\left[1-\gamma(t, t_0)\right]h_{\eta D}$, for instance $h^{(k)}_{AD}(t)\!=\!\left[1-\gamma(t, t_0)\right]h_{AD}$, and $h^{(k)}_{DD}(t)\!=\!\gamma(t, t_0)\Delta_{D}$ with a final potential difference $\Delta_{D}\! =\! 8\bar{t}$ on site $D$.
Although the potential energy on site-$D$ is still finite, we assume that the hopping amplitude decays to zero when the ramping is finished.
The assumption is valid following available experimental results~\cite{Jo:2012bra}.
When ramping to the square lattice, the time-dependent Hamiltonian can be written as
\begin{equation}
	h^{(s)}(\kb)=\left(\begin{array}{cccc}
	0 & h_{AB} & h_{AC} & h^{(s)}_{AD} \\
	h_{AB} & 0 & h^{(s)}_{BC} & h_{BD} \\
	h_{AC} & h^{(s)}_{BC} & 0 & h_{CD} \\
	h^{(s)}_{AD} & h_{BD} & h_{CD} & 0
	\end{array}\right) ,
\end{equation}
where $h^{(s)}_{AD}(t)\!=\!\left[1-\gamma(t, t_0)\right]h_{AD}$ and $h^{(s)}_{BC}(t)\!=\!\left[1-\gamma(t, t_0)\right]h_{BC}$.
Details of the lattice structures are shown in Fig.~\ref{fig:contour}.

\begin{figure}[t]
\begin{center}
\includegraphics[width=0.43\textwidth]{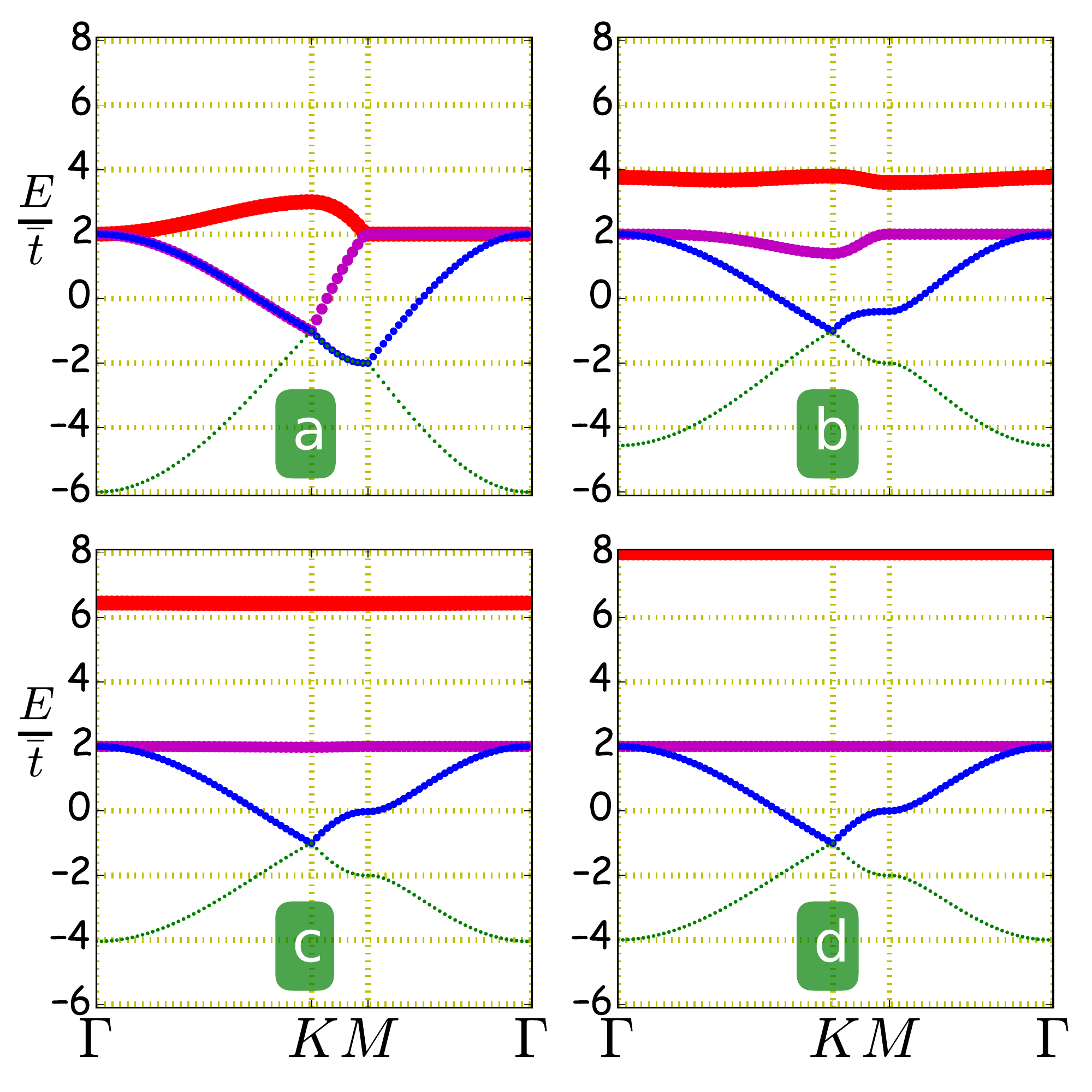}
\caption{(Color online)
Tight-binding bands at different times in a transformation from the triangular to kagome lattice: (a) $t\! =\! 0$, (b) $t\! =\! 0.4t_0$, (c) $t\! =\! 0.8t_0$, and (d) $t\! =\! t_0$.
Here the ramping function is linear with ramping time $t_r\! =\! t_0$.
The third band (purple) develops into a flat band.
}
\label{fig:enk}
\end{center}
\end{figure}

Here we focus on the case of a transformation to the kagome lattice.
The Hamiltonian at each time slot is diagonalized to obtain the corresponding energy spectrum.
In Fig.~\ref{fig:enk}, we plot the energy spectrum at different time slots.
The zone points ($\gamma$, $K$, and $M$) are defined on the kagome Brillouin zone, which is the reason why the spectrum of the triangular lattice shown in Fig.~\ref{fig:enk}(a) has four bands.
As the potential energy ($\Delta_D$) on site-$D$ increases, the forth band energy rises and the third band evolves into a flat band.
After the ramping is completed, the lower three bands forms the kagome band structure and the highest band represents the particles on site-$D$ as shown in Fig.~\ref{fig:enk}(d).

By using the equation of motion of single-particle operators in Heisenberg picture~\cite{mahan2000many}, $i\hbar\partial_{t}c_\mu\! =\! [c_\mu,\mathcal{H}_{tb}]$, the time evolution of the single-particle correlation matrix~\cite{Chien:2012ft} $\mathcal{C}_{\mu,\nu}(\kb)\!=\!\langle c^\dagger_\mu(\kb)c_\nu(\kb)\rangle$ is governed by
\begin{eqnarray}\label{eq:eof}
	&&i\frac{\partial}{\partial t}\langle c^\dagger_\mu(\kb)c_\nu(\kb)\rangle\\
	&=& \sum_{j}[h_{j\mu}(\kb)\langle c^\dagger_j(\kb)c_\nu(\kb)\rangle-h_{\nu j}(\kb)\langle c^\dagger_\mu(\kb)c_j(\kb)\rangle], \nonumber
\end{eqnarray}
which applies to both fermions and bosons.
Here, the summation of $j$ runs over the whole lattice and $\mu,\nu \in\{A,B,C,D\}$.
The initial matrix elements should be determined by the corresponding spin-statistics at given temperature $T$.
For single-component fermions, the filling is given by $N_{tot}/N_{lattice}\!=\!\sum_{\beta, \kb}\langle c^\dagger_\beta(\kb)c_\beta(\kb)\rangle$, where $N_{tot}$ and $N_{lattice}$ denote the total particle number and lattice number, and $\langle c^\dagger_\beta(\kb)c_\beta(\kb)\rangle\!=\!\{\exp((E_{\beta}(\kb)-\mu)/k_BT)+1\}^{-1}$ with $\mu$ and $k_B$ denoting the chemical potential and Boltzmann constant.
The chemical potential is determined from the desired filling fraction and temperature.
Therefore, the initial correlation matrix can be inferred from $\mathcal{C}_{\mu,\nu}(\kb)\!=\!\sum_{\beta, \beta^\prime}(U^\dagger)_{\beta \mu}\langle c^\dagger_\beta(\kb)c_{\beta^\prime}(\kb)\rangle U_{\nu\beta^\prime}$, where the unitary matrix $U$ diagonalizes the Hamiltonian at the given time.

\subsection{Compare to Square Lattice}\label{app:cp}
\begin{figure}[t]
\begin{center}
	\includegraphics[width=0.48\textwidth]{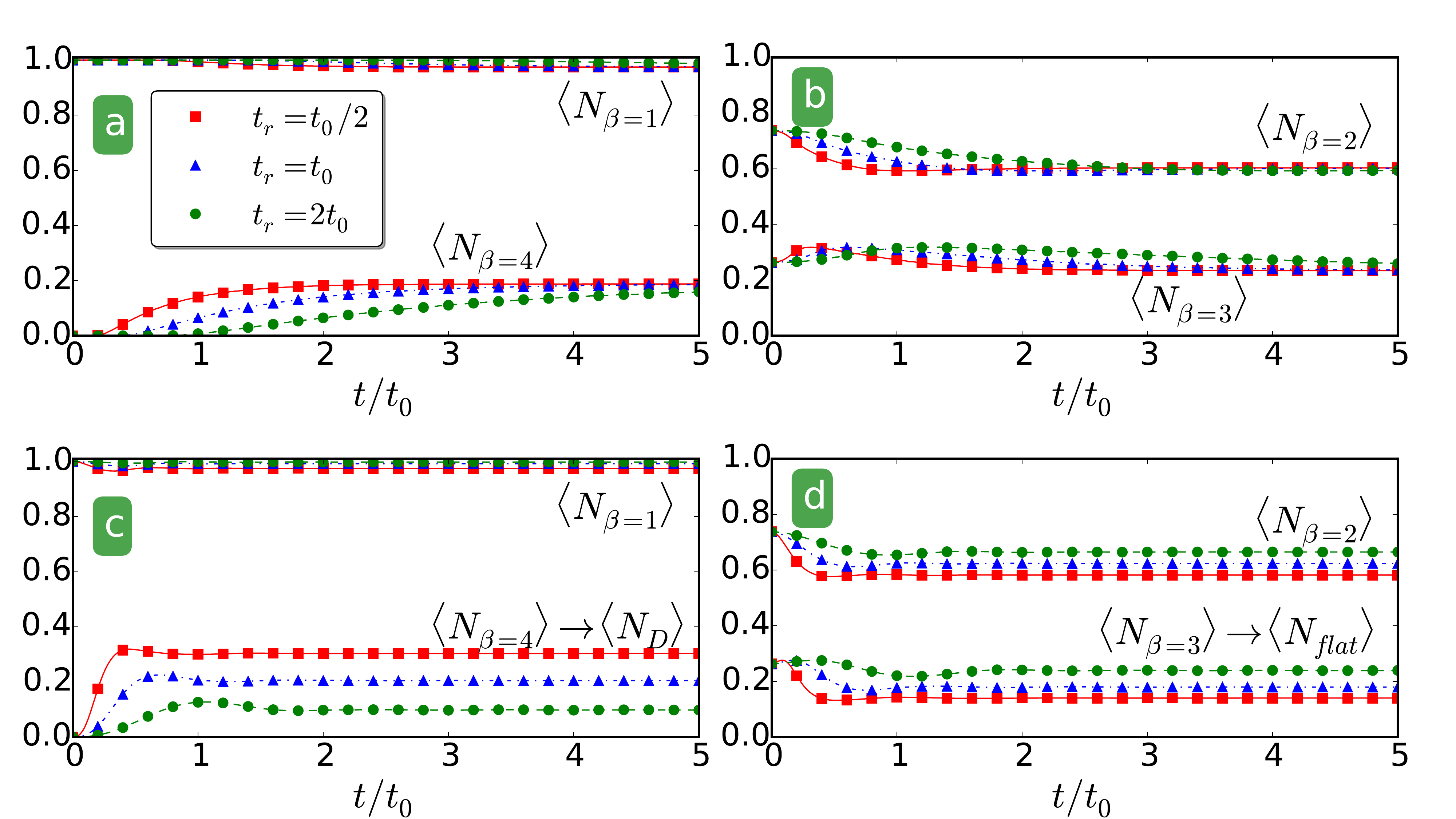}
\caption{(Color online)
Averaged fermion density in each band versus time. (a)-(b) correspond to a transformation into the square lattice and (c)-(d) correspond to a transformation into the kagome lattice at zero temperature.
The different symbols stand for different ramping times.
The filling fraction is set to half-filling.
In (a) and (c), the populations of the lowest ($\beta\!=\!1$) and highest ($\beta\!=\!4$) bands are shown.
In (b) and (d), the populations of the second and third ($\beta\!=\!2,3$) bands are shown.
}
\label{fig:cpsquare}
\end{center}
\end{figure}

In Fig.~\ref{fig:cpsquare}, we show the simulation result by monitoring the dynamics of two different geometrical transformations, one to the kagome lattice and the other to the square lattice.
There is no observable memory effects in the square lattice case and the averaged density on each site approaches the same steady-state value.
The reason we have four bands instead of one for the triangular and square lattices is because we use the same enlarged unit cell to make a fair comparison with the kagome lattice.
On the other hand, memory effect shows up in the kagome lattice as the averaged density in each band varies with different ramping times.
As we mentioned in the main text, a major difference between the square and kagome lattices is the presence of a flat band, which is the source of the memory effects observed here.
We remark that the models used here does not remove the site-$D$ from the lattice but only increased  its on-site energy and set the associated hopping amplitude to zero. Hence we still have residual density on site-$D$ as shown in Fig.~\ref{fig:cpsquare}(c) after a lattice transformation.

\subsection{Non-Thermalized Steady States}\label{app:Thermal}
\begin{figure}[t]
\begin{center}
\includegraphics[width=0.48\textwidth]{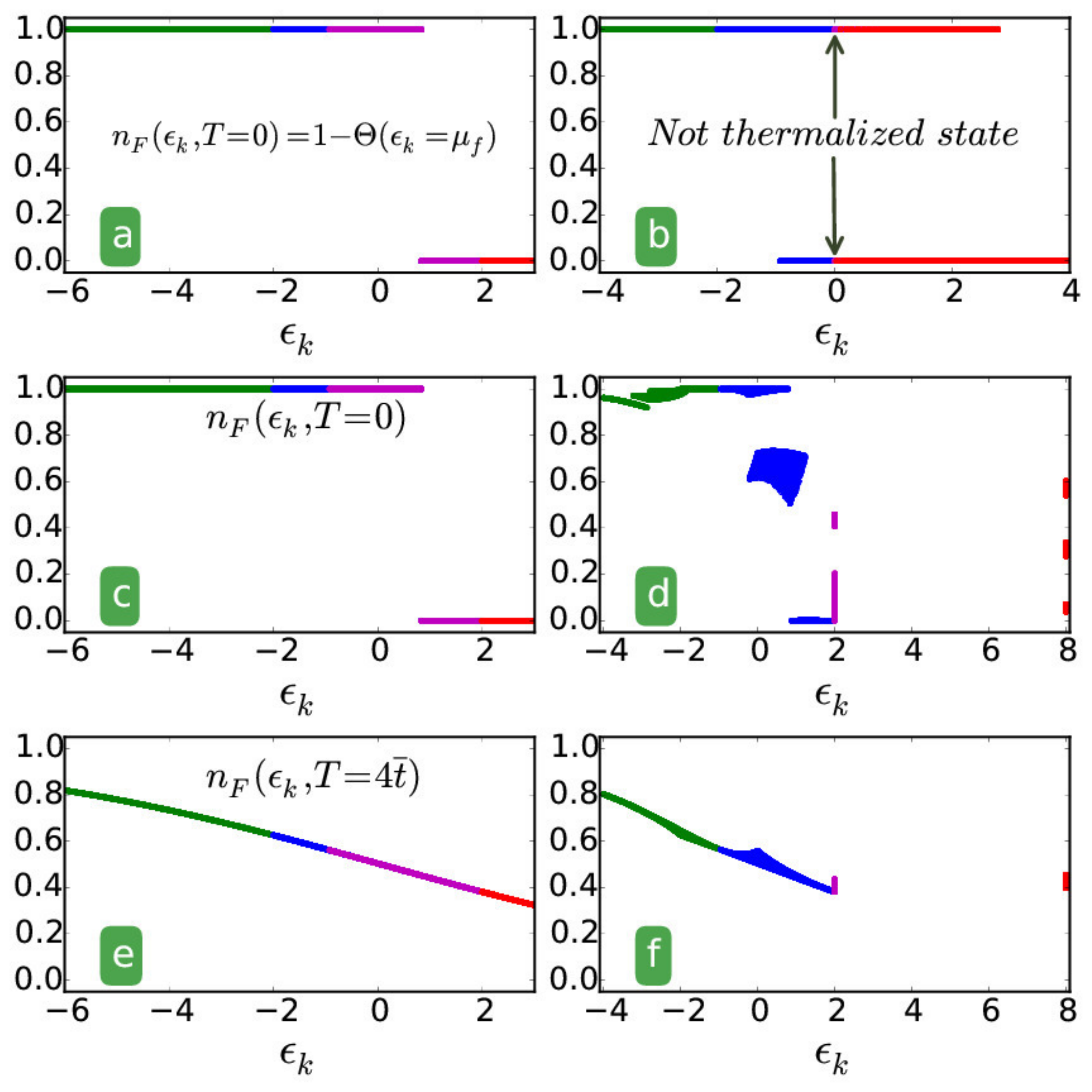}
\caption{(Color online)
Initial $t=0$ (left column) and later $t=6t_0$ (right column) density distributions of fermions for three different cases: (a)-(b) correspond to a transformation to the square lattice at zero temperature.
(c)-(d) correspond to a transformation to the kagome lattice at zero temperature.
(d)-(e) correspond to a transformation to the kagome at $k_B T=4\bar{t}$.
The distributions on the left column are thermal distributions but those on the right are not.
}
\label{fig:thermal}
\end{center}
\end{figure}

In both finite-size and thermodynamic-limit cases, steady states in the long-time limit have been observed.
In a closed system with energy conservation, we found the duration of a quasi-steady state proportional to its size.
We have verified that steady states in the thermodynamic limit observed here do not correspond to any thermal state.
In Fig.~\ref{fig:thermal}, we show three different cases, where the left column shows the initial distributions and the right shows the distributions at a later time after the lattice geometry is transformed.
In the first case, the result of ramping to the square lattice is shown, where non-thermal final distribution supporting a steady state is also observable.
In the second cases, the initial state follows the Fermi-Dirac distribution at zero temperature.
After ramping to the kagome lattice, the later state is not a well-defined thermal distribution since the density distribution depends not only on the energy but also the momentum.
Similar behavior happens when the initial state is a thermalized state in the third case.
In the later state one can see multi-values of the distribution although the distribution is stationary implying a steady state.

\subsection{Filling Dependence}\label{app:FillingBand}
\begin{figure}[t]
\begin{center}
	\includegraphics[width=0.48\textwidth]{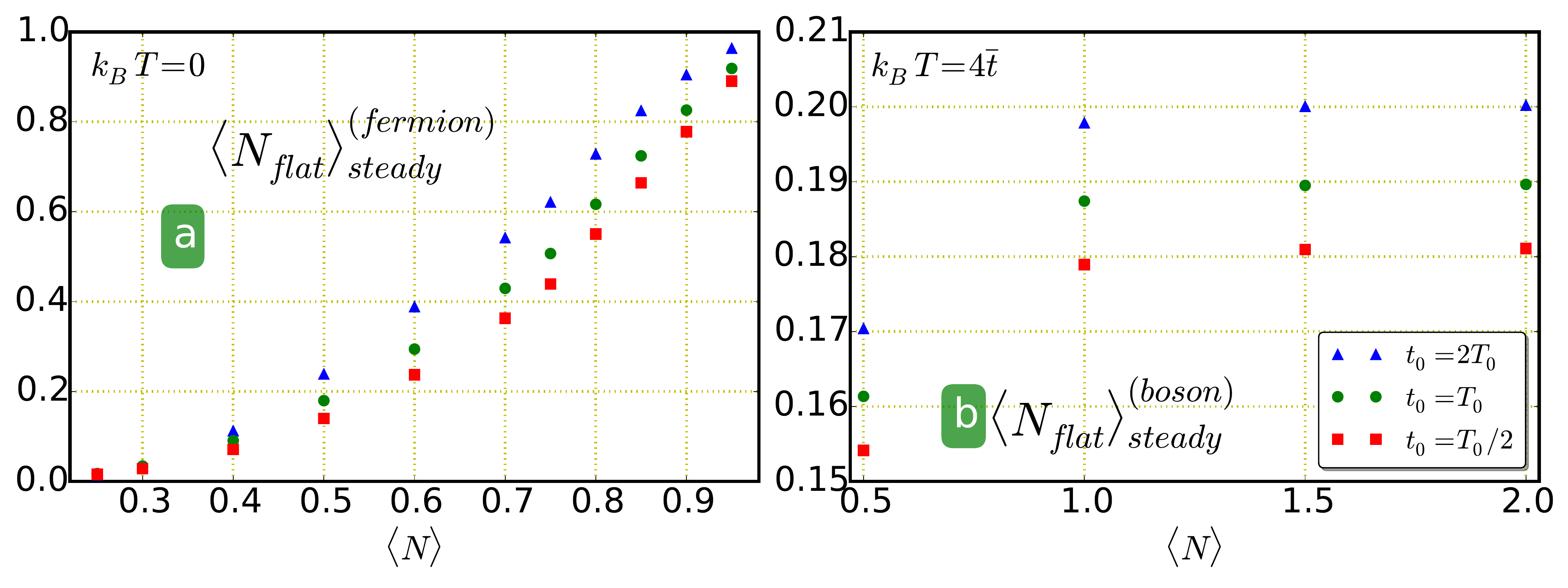}
\caption{(Color online)
(a) Averaged fermion density in a steady state in the kagome flat band versus the initial filling $\langle N\rangle$.
(b) The same plot for non-interacting bosons at $T=4\bar{t}$ with $k_B=1$.
}
\label{fig:kf}
\end{center}
\end{figure}

One subtle difference between the band theory in the thermodynamic limit and the finite-system scheme is that in the band theory only one fermion can occupy one lattice site, while in the finite-size scheme there can be more than one orbitals in a single lattice site because we have used $\Delta^2$ grid points to simulate one single lattice site.
We mentioned that the geometry-induced memory effects are prominent if the initial filling of the triangular lattice ranges from $1/2$ to $3/4$ for fermions at zero temperature.
Fig.~\ref{fig:kf} shows memory effects for single-component fermions at zero temperature and noninteracting bosons at $T=4\bar{t}$.
For fermions, stronger memory effect can be observed around $3/4$ filling since in the initial distribution, the highest band is empty and the third band is completely filled.
Memory effects become weaker once the filling fraction is larger than $3/4$.
When all bands are completely filled, no memory effect can be observed since there is no exchange of populations in the bands.
For bosons, the filling fraction can be any positive number due to the Bose statistics.
In Fig.~\ref{fig:kf}(b), the difference of the flat-band density following different ramping times does not increase significantly as the filling fraction gets larger than one, which suggests that  memory effects of non-interacting bosons would saturate at high density.

\section{Finite Difference Method and Finite Size Effect}\label{app:FiniteDiff}
Here we present the details of our analysis of finite systems.
We use the finite-difference method~\cite{Vudragovic:2012jz,Bao:2006fg} to rewrite continuous derivatives as finite differences defined on discretized square elements of linear size $\Delta$ by choosing $\Delta\!=\!a_L/n_G$, where $n_G$ is the number of grid points in each direction.
Here we discuss finite-size effects and the dependence of our results on $n_G$.
The Schr\"{o}dinger equation governed by Hamiltonian in Eq.~\eqref{eq:H} is discretized on a two dimensional grid labeled by index $j=0, 1, \cdots, n_G$ in each direction.
To be specific, we have the matrix elements (with $\hbar = m = 1$)
\begin{eqnarray}
K_{j,j}&=&\frac{2}{\Delta^2}, \\
K_{j,j+1}
&=&-\frac{1}{2\Delta^2}
\end{eqnarray}
for kinetic energy and
\begin{equation}
V_{jl}\approx \delta_{j,l}V(r_j),
\end{equation}
for potential energy, which is diagonal.
For this method to be valid, $V(r)$ should be slowly varying against $\Delta$.
This approach is also used to discretize the Gross-Pitaevskii equation, in which a non-linear term, $gN_b|\phi|^2$, is included in the equation of motion.

The detailed potential energy for the triangular lattice is
\begin{eqnarray}
	&&V_{tri}(x,y) \\&=& \sin^2(\frac{\pi x}{2a_L})\sin^2(\frac{\pi x}{2a_L}+\frac{\sqrt{3}\pi y}{2a_L})+\sin^2(\frac{\pi x}{2a_L}-\frac{\sqrt{3}\pi y}{2a_L}), \nonumber
\end{eqnarray}
which has a minimum at the origin.
For a transformation to the kagome lattice, another set of lasers focused on the minimum points of $V_{tri}(x,y)$ with twice the wavelength is used, which may be modeled by
\begin{eqnarray}
	&&V_{ramp}^{(k)} \\&=& \sin^2(\frac{\pi x}{a_L})+\sin^2(\frac{\pi x+\sqrt{3}\pi y}{a_L})+\sin^2(\frac{\pi x-\sqrt{3}\pi y}{a_L}).\nonumber
\end{eqnarray}
Finally, to achieve the square lattice, a second set of lasers with shifted focus points and the same wavelength as $V_{tri}(x,y)$ is modeled by
\begin{eqnarray}
	&&V_{ramp}^{(s)} = \sin^2(\frac{x}{2a_L}) \\ &+& \sin^2(\frac{x+\sqrt{3}(y-a_L/2)}{2a_L})+\sin^2(\frac{x-\sqrt{3}(y-a_L/2)}{2a_L}).\nonumber
\end{eqnarray}
The total potential energy is $V(x,y,t)=V_0\left[V_{tri}(x,y)+Z\gamma(t,t_r)V_{ramp}(x,y)\right]$, and we set $V_0=4E_R$ and $Z\!=\!6$ in this work.

\begin{figure*}[t!]
\begin{center}
\includegraphics[width=0.8\textwidth]{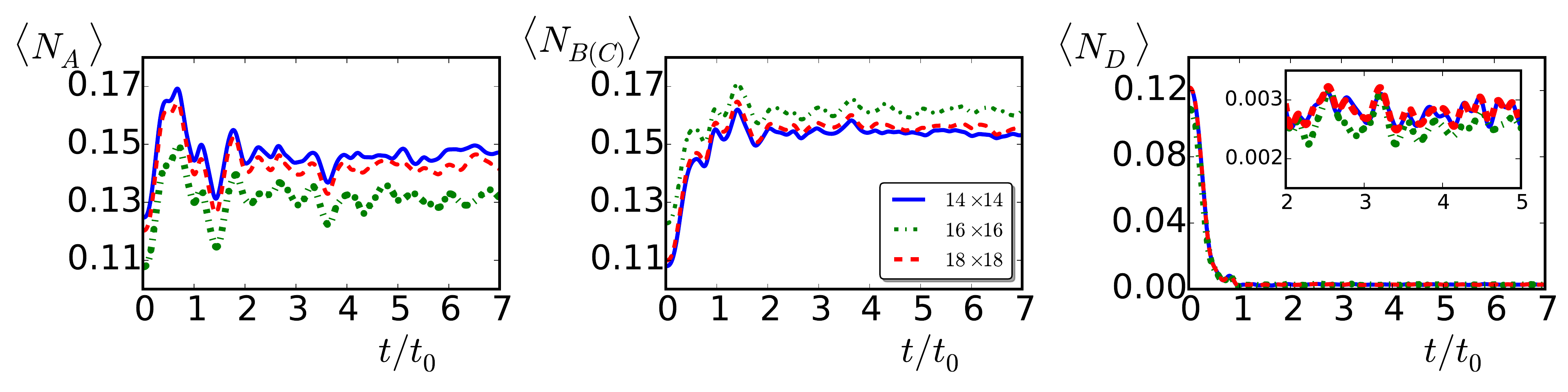}
\includegraphics[width=0.8\textwidth]{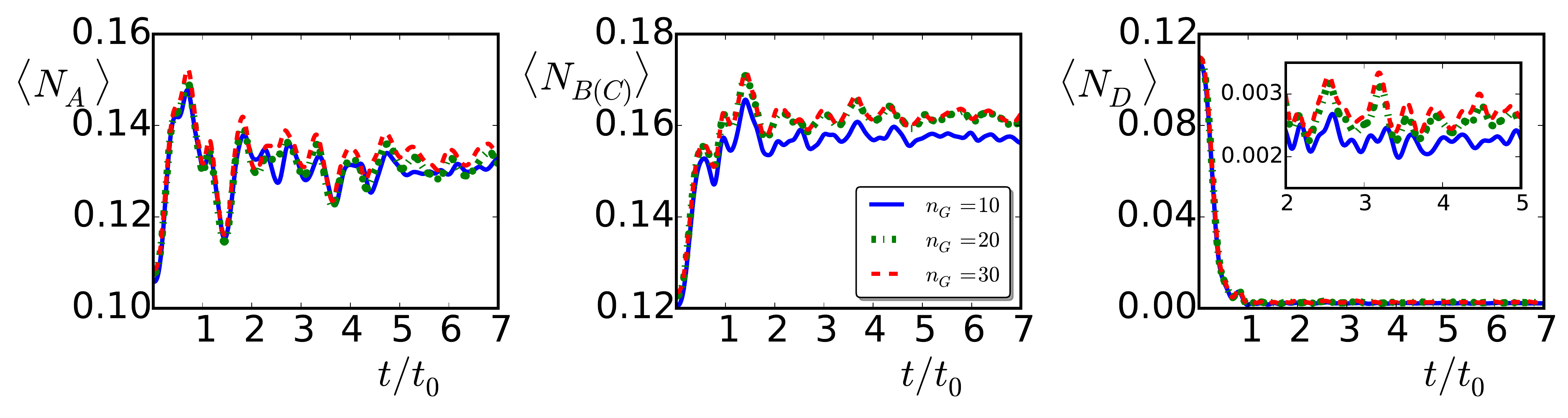}
\caption{(Color online)
Averaged fermion density in a transformation to the kagome lattice with a linear ramping function and ramping time $t_r=t_0$.
The initial filling is set to half-filling.
Top row: Comparisons of different system sizes.
Different colors represent different numbers of effective lattice points on the triangular lattice.
Bottom row: Comparisons of different numbers of grid points ($n_G$).
}
\label{fig:finite}
\end{center}
\end{figure*}

Besides Fig.~\ref{fig:RFKS_NiTime_T0}, we also compare the results with  different system size and different numbers of grid points.
The comparisons are shown in the top (bottom) row of Fig.~\ref{fig:finite} for different numbers of lattice sites (grid points).
For a given system with $(L_x,L_y)\!=\!(8\sqrt{3}, 16)a_L$, the number of minima in each direction is $(N_x, N_y)=(L_x/(\sqrt{3}a_L/2),L_y/a_L)$ and the system has approximately $N_x\times N_y$ lattice sites.
We use the above expression to label system size in Fig.~\ref{fig:finite} and compare different system size with fixed number of grid points $n_G\!=\! 20$.
First, we evaluate the initial densities on the sub-lattices corresponding to the $A, B, C, D$-sites.
Although the initial filling is the same, the density distributions on different sub-lattices are slightly different due to boundary effects because open boundary condition requires wave functions  to vanish at the boundary.
For instance, in a $16\!\times\! 16$ system, site-$A$ and $D$ are located on the boundary, so they have relatively lower density when compared to site $B$ and $C$.
On the other hand, in $18\!\times\! 18$ and $14\! \times\! 14$ systems the densities on site-$A$ and $D$ are closed to each other.
By considering the small offsets from the initial densities on different sites, the results from different system sizes agree qualitatively.
Most importantly, the residual density on site-$D$ is insensitive of system size and thus can serve as a signature of memory effect.
For different numbers of grid points, we choose the system size to be $16\times16$.
As mentioned before, the number $n_G$ needs to be large enough to make the above approach valid, a condition equivalent to the requirement that $V(x)$ should be slow vary against $\Delta$.
Moreover, the choice of $n_G$ may be interpreted as the number of effective orbitals on each lattice site. A finite number of effective orbitals allows a more detailed description of finite-size systems and rules out a dynamically-generated insulator~\cite{Chien:2013ef}, .

The results with three different values $n_G$ are shown in the bottom row of Fig.~\ref{fig:finite}. The results with $n_G\!=\! 20$ and $30$ agree with each other quite well.
For the $n_G\!=\! 10$ case, the initial density distribution is slightly different due to boundary effects, especially on site-$B$ and site-$C$, leading to a slightly different quasi-steady state value.
Nevertheless, all cases show the same qualitative behavior.


\begin{thebibliography}{70}%
\makeatletter
\providecommand \@ifxundefined [1]{%
 \@ifx{#1\undefined}
}%
\providecommand \@ifnum [1]{%
 \ifnum #1\expandafter \@firstoftwo
 \else \expandafter \@secondoftwo
 \fi
}%
\providecommand \@ifx [1]{%
 \ifx #1\expandafter \@firstoftwo
 \else \expandafter \@secondoftwo
 \fi
}%
\providecommand \natexlab [1]{#1}%
\providecommand \enquote  [1]{``#1''}%
\providecommand \bibnamefont  [1]{#1}%
\providecommand \bibfnamefont [1]{#1}%
\providecommand \citenamefont [1]{#1}%
\providecommand \href@noop [0]{\@secondoftwo}%
\providecommand \href [0]{\begingroup \@sanitize@url \@href}%
\providecommand \@href[1]{\@@startlink{#1}\@@href}%
\providecommand \@@href[1]{\endgroup#1\@@endlink}%
\providecommand \@sanitize@url [0]{\catcode `\\12\catcode `\$12\catcode
  `\&12\catcode `\#12\catcode `\^12\catcode `\_12\catcode `\%12\relax}%
\providecommand \@@startlink[1]{}%
\providecommand \@@endlink[0]{}%
\providecommand \url  [0]{\begingroup\@sanitize@url \@url }%
\providecommand \@url [1]{\endgroup\@href {#1}{\urlprefix }}%
\providecommand \urlprefix  [0]{URL }%
\providecommand \Eprint [0]{\href }%
\providecommand \doibase [0]{http://dx.doi.org/}%
\providecommand \selectlanguage [0]{\@gobble}%
\providecommand \bibinfo  [0]{\@secondoftwo}%
\providecommand \bibfield  [0]{\@secondoftwo}%
\providecommand \translation [1]{[#1]}%
\providecommand \BibitemOpen [0]{}%
\providecommand \bibitemStop [0]{}%
\providecommand \bibitemNoStop [0]{.\EOS\space}%
\providecommand \EOS [0]{\spacefactor3000\relax}%
\providecommand \BibitemShut  [1]{\csname bibitem#1\endcsname}%
\let\auto@bib@innerbib\@empty
\bibitem [{\citenamefont {Jackson}(1999)}]{JacksonBook}%
  \BibitemOpen
  \bibfield  {author} {\bibinfo {author} {\bibfnamefont {J.~D.}\ \bibnamefont
  {Jackson}},\ }\href@noop {} {\emph {\bibinfo {title} {Classical
  electrodynamics}}},\ \bibinfo {edition} {3rd}\ ed.\ (\bibinfo  {publisher}
  {John Wiley and Sons},\ \bibinfo {year} {1999})\BibitemShut {NoStop}%
\bibitem [{\citenamefont {Sun}\ \emph {et~al.}(2003)\citenamefont {Sun},
  \citenamefont {Salamon}, \citenamefont {Garnier},\ and\ \citenamefont
  {Averback}}]{Sun:2003iz}%
  \BibitemOpen
  \bibfield  {author} {\bibinfo {author} {\bibfnamefont {Y.}~\bibnamefont
  {Sun}}, \bibinfo {author} {\bibfnamefont {M.~B.}\ \bibnamefont {Salamon}},
  \bibinfo {author} {\bibfnamefont {K.}~\bibnamefont {Garnier}}, \ and\
  \bibinfo {author} {\bibfnamefont {R.~S.}\ \bibnamefont {Averback}},\
  }\bibfield  {title} {\enquote {\bibinfo {title} {{Memory Effects in an
  Interacting Magnetic Nanoparticle System}},}\ }\href@noop {} {\bibfield
  {journal} {\bibinfo  {journal} {Phys. Rev. Lett.}\ }\textbf {\bibinfo
  {volume} {91}},\ \bibinfo {pages} {167206} (\bibinfo {year}
  {2003})}\BibitemShut {NoStop}%
\bibitem [{\citenamefont {Liden}\ and\ \citenamefont
  {Reddy}(2001)}]{LidenBook}%
  \BibitemOpen
  \bibfield  {author} {\bibinfo {author} {\bibfnamefont {D.}~\bibnamefont
  {Liden}}\ and\ \bibinfo {author} {\bibfnamefont {T.~B.}\ \bibnamefont
  {Reddy}},\ }\href@noop {} {\emph {\bibinfo {title} {Handbook of
  batteries}}},\ \bibinfo {edition} {3rd}\ ed.\ (\bibinfo  {publisher}
  {McGraw-Hill},\ \bibinfo {year} {2001})\BibitemShut {NoStop}%
\bibitem [{\citenamefont {Sasaki}\ \emph {et~al.}(2013)\citenamefont {Sasaki},
  \citenamefont {Ukyo},\ and\ \citenamefont {Nov{\'a}k}}]{Sasaki:2013cd}%
  \BibitemOpen
  \bibfield  {author} {\bibinfo {author} {\bibfnamefont {T.}~\bibnamefont
  {Sasaki}}, \bibinfo {author} {\bibfnamefont {Y.}~\bibnamefont {Ukyo}}, \ and\
  \bibinfo {author} {\bibfnamefont {P.}~\bibnamefont {Nov{\'a}k}},\ }\bibfield
  {title} {\enquote {\bibinfo {title} {{Memory effect in a lithium-ion
  battery}},}\ }\href@noop {} {\bibfield  {journal} {\bibinfo  {journal}
  {Nature Materials}\ }\textbf {\bibinfo {volume} {12}},\ \bibinfo {pages}
  {569--575} (\bibinfo {year} {2013})}\BibitemShut {NoStop}%
\bibitem [{\citenamefont {Gershenfeld}(2000)}]{gershenfeld2000physics}%
  \BibitemOpen
  \bibfield  {author} {\bibinfo {author} {\bibfnamefont {N.~A.}\ \bibnamefont
  {Gershenfeld}},\ }\href@noop {} {\emph {\bibinfo {title} {The physics of
  information technology}}},\ Vol.~\bibinfo {volume} {25}\ (\bibinfo
  {publisher} {Cambridge University Press Cambridge},\ \bibinfo {year}
  {2000})\BibitemShut {NoStop}%
\bibitem [{\citenamefont {Ebara}\ \emph {et~al.}(2014)\citenamefont {Ebara},
  \citenamefont {Kotsuchibashi}, \citenamefont {Narain}, \citenamefont {Idota},
  \citenamefont {Kim}, \citenamefont {Hoffman}, \citenamefont {Uto},\ and\
  \citenamefont {Aoyagi}}]{EbaraBook}%
  \BibitemOpen
  \bibfield  {author} {\bibinfo {author} {\bibfnamefont {M.}~\bibnamefont
  {Ebara}}, \bibinfo {author} {\bibfnamefont {Y.}~\bibnamefont
  {Kotsuchibashi}}, \bibinfo {author} {\bibfnamefont {R.}~\bibnamefont
  {Narain}}, \bibinfo {author} {\bibfnamefont {N.}~\bibnamefont {Idota}},
  \bibinfo {author} {\bibfnamefont {Y.~J.}\ \bibnamefont {Kim}}, \bibinfo
  {author} {\bibfnamefont {J.~M.}\ \bibnamefont {Hoffman}}, \bibinfo {author}
  {\bibfnamefont {K.}~\bibnamefont {Uto}}, \ and\ \bibinfo {author}
  {\bibfnamefont {T.}~\bibnamefont {Aoyagi}},\ }\href@noop {} {\emph {\bibinfo
  {title} {Smart biomaterials}}}\ (\bibinfo  {publisher} {Springer},\ \bibinfo
  {address} {Berlin},\ \bibinfo {year} {2014})\BibitemShut {NoStop}%
\bibitem [{\citenamefont {Cui}(2014)}]{CuiSMA}%
  \BibitemOpen
  \bibfield  {author} {\bibinfo {author} {\bibfnamefont {J.}~\bibnamefont
  {Cui}},\ }\bibfield  {title} {\enquote {\bibinfo {title} {Shape memory alloys
  and their applications in power generation and refrigeration},}\ }in\
  \href@noop {} {\emph {\bibinfo {booktitle} {Mesoscopi phenomena in
  multifunctional materials: Synthesis, characterization, modeling, and
  applications}}},\ \bibinfo {editor} {edited by\ \bibinfo {editor}
  {\bibfnamefont {A.}~\bibnamefont {Saxena}}\ and\ \bibinfo {editor}
  {\bibfnamefont {A.}~\bibnamefont {Planes}}}\ (\bibinfo  {publisher}
  {Springer-Verlag},\ \bibinfo {address} {Berlin},\ \bibinfo {year}
  {2014})\BibitemShut {NoStop}%
\bibitem [{\citenamefont {Pershin}\ and\ \citenamefont
  {Di~Ventra}(2011)}]{Pershin:2011ie}%
  \BibitemOpen
  \bibfield  {author} {\bibinfo {author} {\bibfnamefont {Y.~V.}\ \bibnamefont
  {Pershin}}\ and\ \bibinfo {author} {\bibfnamefont {M.}~\bibnamefont
  {Di~Ventra}},\ }\bibfield  {title} {\enquote {\bibinfo {title} {Memory
  effects in complex materials and nanoscale systems},}\ }\href
  {http://dx.doi.org/10.1080/00018732.2010.544961} {\bibfield  {journal}
  {\bibinfo  {journal} {Adv. Phys.}\ }\textbf {\bibinfo {volume} {60}},\
  \bibinfo {pages} {145--227} (\bibinfo {year} {2011})}\BibitemShut {NoStop}%
\bibitem [{\citenamefont {Marani}\ \emph {et~al.}(2015)\citenamefont {Marani},
  \citenamefont {Gelao},\ and\ \citenamefont {Perri}}]{Marani15}%
  \BibitemOpen
  \bibfield  {author} {\bibinfo {author} {\bibfnamefont {R.}~\bibnamefont
  {Marani}}, \bibinfo {author} {\bibfnamefont {G.}~\bibnamefont {Gelao}}, \
  and\ \bibinfo {author} {\bibfnamefont {A.~G.}\ \bibnamefont {Perri}},\
  }\bibfield  {title} {\enquote {\bibinfo {title} {A review on memristor
  applications},}\ }\href@noop {} {\bibfield  {journal} {\bibinfo  {journal}
  {arXiv:1506.06899}\ } (\bibinfo {year} {2015})}\BibitemShut {NoStop}%
\bibitem [{\citenamefont {Gilbert}\ \emph {et~al.}(2015)\citenamefont
  {Gilbert}, \citenamefont {Chern}, \citenamefont {Fore}, \citenamefont {Lao},
  \citenamefont {Zhang}, \citenamefont {Nisoli},\ and\ \citenamefont
  {Schiffer}}]{Gilbert:2015vx}%
  \BibitemOpen
  \bibfield  {author} {\bibinfo {author} {\bibfnamefont {I.}~\bibnamefont
  {Gilbert}}, \bibinfo {author} {\bibfnamefont {G.-W.}\ \bibnamefont {Chern}},
  \bibinfo {author} {\bibfnamefont {B.}~\bibnamefont {Fore}}, \bibinfo {author}
  {\bibfnamefont {Y.}~\bibnamefont {Lao}}, \bibinfo {author} {\bibfnamefont
  {S.}~\bibnamefont {Zhang}}, \bibinfo {author} {\bibfnamefont
  {C.}~\bibnamefont {Nisoli}}, \ and\ \bibinfo {author} {\bibfnamefont
  {P.}~\bibnamefont {Schiffer}},\ }\bibfield  {title} {\enquote {\bibinfo
  {title} {Direct visualization of memory effects in artificial spin ice},}\
  }\href@noop {} {\bibfield  {journal} {\bibinfo  {journal} {Phys. Rev. B}\
  }\textbf {\bibinfo {volume} {92}},\ \bibinfo {pages} {104417} (\bibinfo
  {year} {2015})}\BibitemShut {NoStop}%
\bibitem [{\citenamefont {Ashcroft}\ and\ \citenamefont
  {Mermin}(1976)}]{ashcroft1976solid}%
  \BibitemOpen
  \bibfield  {author} {\bibinfo {author} {\bibfnamefont {N.~W.}\ \bibnamefont
  {Ashcroft}}\ and\ \bibinfo {author} {\bibfnamefont {N.~D.}\ \bibnamefont
  {Mermin}},\ }\href@noop {} {\emph {\bibinfo {title} {{Solid state
  physics}}}}\ (\bibinfo  {publisher} {Saunders College},\ \bibinfo {year}
  {1976})\BibitemShut {NoStop}%
\bibitem [{\citenamefont {Chien}\ and\ \citenamefont
  {Di~Ventra}(2013)}]{ChienPS13}%
  \BibitemOpen
  \bibfield  {author} {\bibinfo {author} {\bibfnamefont {C.~C.}\ \bibnamefont
  {Chien}}\ and\ \bibinfo {author} {\bibfnamefont {M.}~\bibnamefont
  {Di~Ventra}},\ }\bibfield  {title} {\enquote {\bibinfo {title} {Controlling
  transport of ultra-cold atoms in 1d optical lattices with artificial gauge
  fields},}\ }\href@noop {} {\bibfield  {journal} {\bibinfo  {journal} {Phys.
  Rev. A}\ }\textbf {\bibinfo {volume} {87}},\ \bibinfo {pages} {023609}
  (\bibinfo {year} {2013})}\BibitemShut {NoStop}%
\bibitem [{\citenamefont {Cornean}\ \emph {et~al.}(2013)\citenamefont
  {Cornean}, \citenamefont {Jensen},\ and\ \citenamefont
  {Nenciu}}]{Cornean:2013kz}%
  \BibitemOpen
  \bibfield  {author} {\bibinfo {author} {\bibfnamefont {H.~D.}\ \bibnamefont
  {Cornean}}, \bibinfo {author} {\bibfnamefont {A.}~\bibnamefont {Jensen}}, \
  and\ \bibinfo {author} {\bibfnamefont {G.}~\bibnamefont {Nenciu}},\
  }\bibfield  {title} {\enquote {\bibinfo {title} {{Memory Effects in
  Non-Interacting Mesoscopic Transport}},}\ }\href@noop {} {\bibfield
  {journal} {\bibinfo  {journal} {Ann. Henri Poincar{\'e}}\ }\textbf {\bibinfo
  {volume} {15}},\ \bibinfo {pages} {1919--1943} (\bibinfo {year}
  {2013})}\BibitemShut {NoStop}%
\bibitem [{\citenamefont {Bloch}\ and\ \citenamefont
  {Zwerger}(2008)}]{Bloch:2008gl}%
  \BibitemOpen
  \bibfield  {author} {\bibinfo {author} {\bibfnamefont {I.}~\bibnamefont
  {Bloch}}\ and\ \bibinfo {author} {\bibfnamefont {W.}~\bibnamefont
  {Zwerger}},\ }\bibfield  {title} {\enquote {\bibinfo {title} {{Many-body
  physics with ultracold gases}},}\ }\href@noop {} {\bibfield  {journal}
  {\bibinfo  {journal} {Rev. Mod. Phys.}\ }\textbf {\bibinfo {volume} {80}},\
  \bibinfo {pages} {885--964} (\bibinfo {year} {2008})}\BibitemShut {NoStop}%
\bibitem [{\citenamefont {Pethick}\ and\ \citenamefont
  {Smith}(2010)}]{Pethick:2010gy}%
  \BibitemOpen
  \bibfield  {author} {\bibinfo {author} {\bibfnamefont {C.~J.}\ \bibnamefont
  {Pethick}}\ and\ \bibinfo {author} {\bibfnamefont {H.}~\bibnamefont
  {Smith}},\ }\href@noop {} {\emph {\bibinfo {title}
  {{Bose{\textendash}Einstein Condensation in Dilute Gases}}}},\ \bibinfo
  {edition} {2nd}\ ed.\ (\bibinfo  {publisher} {Cambridge University Press},\
  \bibinfo {address} {Cambridge},\ \bibinfo {year} {2010})\BibitemShut
  {NoStop}%
\bibitem [{\citenamefont {Stoof}\ \emph {et~al.}(2008)\citenamefont {Stoof},
  \citenamefont {Dickerscheid},\ and\ \citenamefont {Gubbels}}]{Stoof:2008ho}%
  \BibitemOpen
  \bibfield  {author} {\bibinfo {author} {\bibfnamefont {H.~T.~C.}\
  \bibnamefont {Stoof}}, \bibinfo {author} {\bibfnamefont {D.~B.~M.}\
  \bibnamefont {Dickerscheid}}, \ and\ \bibinfo {author} {\bibfnamefont
  {K.}~\bibnamefont {Gubbels}},\ }\href@noop {} {\emph {\bibinfo {title}
  {{Ultracold Quantum Fields}}}},\ Theoretical and Mathematical Physics\
  (\bibinfo  {publisher} {Springer Netherlands},\ \bibinfo {address}
  {Dordrecht},\ \bibinfo {year} {2008})\BibitemShut {NoStop}%
\bibitem [{\citenamefont {Chin}\ \emph {et~al.}(2010)\citenamefont {Chin},
  \citenamefont {Grimm}, \citenamefont {Julienne},\ and\ \citenamefont
  {Tiesinga}}]{Chin:2010kl}%
  \BibitemOpen
  \bibfield  {author} {\bibinfo {author} {\bibfnamefont {C.}~\bibnamefont
  {Chin}}, \bibinfo {author} {\bibfnamefont {R.}~\bibnamefont {Grimm}},
  \bibinfo {author} {\bibfnamefont {P.~S.}\ \bibnamefont {Julienne}}, \ and\
  \bibinfo {author} {\bibfnamefont {E.}~\bibnamefont {Tiesinga}},\ }\bibfield
  {title} {\enquote {\bibinfo {title} {{Feshbach resonances in ultracold
  gases}},}\ }\href@noop {} {\bibfield  {journal} {\bibinfo  {journal} {Rev.
  Mod. Phys.}\ }\textbf {\bibinfo {volume} {82}},\ \bibinfo {pages}
  {1225--1286} (\bibinfo {year} {2010})}\BibitemShut {NoStop}%
\bibitem [{\citenamefont {Eckel}\ \emph {et~al.}(2014)\citenamefont {Eckel},
  \citenamefont {Lee}, \citenamefont {Jendrzejewski}, \citenamefont {Murray},
  \citenamefont {Clark}, \citenamefont {Lobb}, \citenamefont {Phillips},
  \citenamefont {Edwards},\ and\ \citenamefont {Campbell}}]{Eckel:2014gf}%
  \BibitemOpen
  \bibfield  {author} {\bibinfo {author} {\bibfnamefont {S.}~\bibnamefont
  {Eckel}}, \bibinfo {author} {\bibfnamefont {J.~G.}\ \bibnamefont {Lee}},
  \bibinfo {author} {\bibfnamefont {F.}~\bibnamefont {Jendrzejewski}}, \bibinfo
  {author} {\bibfnamefont {N.}~\bibnamefont {Murray}}, \bibinfo {author}
  {\bibfnamefont {C.~W.}\ \bibnamefont {Clark}}, \bibinfo {author}
  {\bibfnamefont {C.~J.}\ \bibnamefont {Lobb}}, \bibinfo {author}
  {\bibfnamefont {W.~D.}\ \bibnamefont {Phillips}}, \bibinfo {author}
  {\bibfnamefont {M.}~\bibnamefont {Edwards}}, \ and\ \bibinfo {author}
  {\bibfnamefont {G.~K.}\ \bibnamefont {Campbell}},\ }\bibfield  {title}
  {\enquote {\bibinfo {title} {{Hysteresis in a quantized superfluid
  `atomtronic' circuit}},}\ }\href@noop {} {\bibfield  {journal} {\bibinfo
  {journal} {Nature}\ }\textbf {\bibinfo {volume} {506}},\ \bibinfo {pages}
  {200--203} (\bibinfo {year} {2014})}\BibitemShut {NoStop}%
\bibitem [{\citenamefont {Chien}\ \emph {et~al.}(2015)\citenamefont {Chien},
  \citenamefont {Peotta},\ and\ \citenamefont {Di~Ventra}}]{ChienTranreview}%
  \BibitemOpen
  \bibfield  {author} {\bibinfo {author} {\bibfnamefont {C.~C.}\ \bibnamefont
  {Chien}}, \bibinfo {author} {\bibfnamefont {S.}~\bibnamefont {Peotta}}, \
  and\ \bibinfo {author} {\bibfnamefont {M.}~\bibnamefont {Di~Ventra}},\
  }\bibfield  {title} {\enquote {\bibinfo {title} {Quantum transport in
  ultracold atoms},}\ }\href@noop {} {\bibfield  {journal} {\bibinfo  {journal}
  {Nat. Phys.}\ }\textbf {\bibinfo {volume} {11}},\ \bibinfo {pages} {998}
  (\bibinfo {year} {2015})}\BibitemShut {NoStop}%
\bibitem [{\citenamefont {Salger}\ \emph {et~al.}(2009)\citenamefont {Salger},
  \citenamefont {Kling}, \citenamefont {Hecking}, \citenamefont {Geckeler},
  \citenamefont {Morales-Molina},\ and\ \citenamefont {Weitz}}]{Salger:2009ci}%
  \BibitemOpen
  \bibfield  {author} {\bibinfo {author} {\bibfnamefont {T.}~\bibnamefont
  {Salger}}, \bibinfo {author} {\bibfnamefont {S.}~\bibnamefont {Kling}},
  \bibinfo {author} {\bibfnamefont {T.}~\bibnamefont {Hecking}}, \bibinfo
  {author} {\bibfnamefont {C.}~\bibnamefont {Geckeler}}, \bibinfo {author}
  {\bibfnamefont {L.}~\bibnamefont {Morales-Molina}}, \ and\ \bibinfo {author}
  {\bibfnamefont {M.}~\bibnamefont {Weitz}},\ }\bibfield  {title} {\enquote
  {\bibinfo {title} {{Directed Transport of Atoms in a Hamiltonian Quantum
  Ratchet}},}\ }\href@noop {} {\bibfield  {journal} {\bibinfo  {journal}
  {Science}\ }\textbf {\bibinfo {volume} {326}},\ \bibinfo {pages} {1241--1243}
  (\bibinfo {year} {2009})}\BibitemShut {NoStop}%
\bibitem [{\citenamefont {Schneider}\ \emph {et~al.}(2012)\citenamefont
  {Schneider}, \citenamefont {Hackerm{\"u}ller}, \citenamefont {Ronzheimer},
  \citenamefont {Will}, \citenamefont {Braun}, \citenamefont {Best},
  \citenamefont {Bloch}, \citenamefont {Demler}, \citenamefont {Mandt},
  \citenamefont {Rasch},\ and\ \citenamefont {Rosch}}]{Schneider:2012ke}%
  \BibitemOpen
  \bibfield  {author} {\bibinfo {author} {\bibfnamefont {U.}~\bibnamefont
  {Schneider}}, \bibinfo {author} {\bibfnamefont {L.}~\bibnamefont
  {Hackerm{\"u}ller}}, \bibinfo {author} {\bibfnamefont {J.~P.}\ \bibnamefont
  {Ronzheimer}}, \bibinfo {author} {\bibfnamefont {S.}~\bibnamefont {Will}},
  \bibinfo {author} {\bibfnamefont {S.}~\bibnamefont {Braun}}, \bibinfo
  {author} {\bibfnamefont {T.}~\bibnamefont {Best}}, \bibinfo {author}
  {\bibfnamefont {I.}~\bibnamefont {Bloch}}, \bibinfo {author} {\bibfnamefont
  {E.}~\bibnamefont {Demler}}, \bibinfo {author} {\bibfnamefont
  {S.}~\bibnamefont {Mandt}}, \bibinfo {author} {\bibfnamefont
  {D.}~\bibnamefont {Rasch}}, \ and\ \bibinfo {author} {\bibfnamefont
  {A.}~\bibnamefont {Rosch}},\ }\bibfield  {title} {\enquote {\bibinfo {title}
  {{Fermionic transport and out-of-equilibrium dynamics in a homogeneous
  Hubbard model with ultracold atoms}},}\ }\href@noop {} {\bibfield  {journal}
  {\bibinfo  {journal} {Nat. Phys.}\ }\textbf {\bibinfo {volume} {8}},\
  \bibinfo {pages} {213--218} (\bibinfo {year} {2012})}\BibitemShut {NoStop}%
\bibitem [{\citenamefont {Ott}\ \emph {et~al.}(2004)\citenamefont {Ott},
  \citenamefont {de~Mirandes}, \citenamefont {Ferlaino}, \citenamefont {Roati},
  \citenamefont {Modugno},\ and\ \citenamefont {Inguscio}}]{Ott:2004jk}%
  \BibitemOpen
  \bibfield  {author} {\bibinfo {author} {\bibfnamefont {H.}~\bibnamefont
  {Ott}}, \bibinfo {author} {\bibfnamefont {E.}~\bibnamefont {de~Mirandes}},
  \bibinfo {author} {\bibfnamefont {F.}~\bibnamefont {Ferlaino}}, \bibinfo
  {author} {\bibfnamefont {G.}~\bibnamefont {Roati}}, \bibinfo {author}
  {\bibfnamefont {G.}~\bibnamefont {Modugno}}, \ and\ \bibinfo {author}
  {\bibfnamefont {M.}~\bibnamefont {Inguscio}},\ }\bibfield  {title} {\enquote
  {\bibinfo {title} {{Collisionally Induced Transport in Periodic
  Potentials}},}\ }\href@noop {} {\bibfield  {journal} {\bibinfo  {journal}
  {Phys. Rev. Lett.}\ }\textbf {\bibinfo {volume} {92}},\ \bibinfo {pages}
  {160601} (\bibinfo {year} {2004})}\BibitemShut {NoStop}%
\bibitem [{\citenamefont {Partridge}\ \emph {et~al.}(2006)\citenamefont
  {Partridge}, \citenamefont {Li}, \citenamefont {Liao}, \citenamefont {Hulet},
  \citenamefont {Haque},\ and\ \citenamefont {Stoof}}]{Partridge:2006fq}%
  \BibitemOpen
  \bibfield  {author} {\bibinfo {author} {\bibfnamefont {G.~B.}\ \bibnamefont
  {Partridge}}, \bibinfo {author} {\bibfnamefont {W.}~\bibnamefont {Li}},
  \bibinfo {author} {\bibfnamefont {Y.~A.}\ \bibnamefont {Liao}}, \bibinfo
  {author} {\bibfnamefont {R.~G.}\ \bibnamefont {Hulet}}, \bibinfo {author}
  {\bibfnamefont {M.}~\bibnamefont {Haque}}, \ and\ \bibinfo {author}
  {\bibfnamefont {H.~T.~C.}~\bibnamefont {Stoof}},\ }\bibfield  {title} {\enquote
  {\bibinfo {title} {{Deformation of a Trapped Fermi Gas with Unequal Spin
  Populations}},}\ }\href@noop {} {\bibfield  {journal} {\bibinfo  {journal}
  {Phys. Rev. Lett.}\ }\textbf {\bibinfo {volume} {97}},\ \bibinfo {pages}
  {190407} (\bibinfo {year} {2006})}\BibitemShut {NoStop}%
\bibitem [{\citenamefont {Zwierlein}\ \emph {et~al.}(2006)\citenamefont
  {Zwierlein}, \citenamefont {Schirotzek}, \citenamefont {Schunck},\ and\
  \citenamefont {Ketterle}}]{Zwierlein:2006gb}%
  \BibitemOpen
  \bibfield  {author} {\bibinfo {author} {\bibfnamefont {M.~W.}\ \bibnamefont
  {Zwierlein}}, \bibinfo {author} {\bibfnamefont {A.}~\bibnamefont
  {Schirotzek}}, \bibinfo {author} {\bibfnamefont {C.~H.}\ \bibnamefont
  {Schunck}}, \ and\ \bibinfo {author} {\bibfnamefont {W.}~\bibnamefont
  {Ketterle}},\ }\bibfield  {title} {\enquote {\bibinfo {title} {{Fermionic
  Superfluidity with Imbalanced Spin Populations}},}\ }\href@noop {} {\bibfield
   {journal} {\bibinfo  {journal} {Science}\ }\textbf {\bibinfo {volume}
  {311}},\ \bibinfo {pages} {492--496} (\bibinfo {year} {2006})}\BibitemShut
  {NoStop}%
\bibitem [{\citenamefont {Fukuhara}\ \emph {et~al.}(2009)\citenamefont
  {Fukuhara}, \citenamefont {Sugawa}, \citenamefont {Takasu},\ and\
  \citenamefont {Takahashi}}]{Fukuhara:2009bv}%
  \BibitemOpen
  \bibfield  {author} {\bibinfo {author} {\bibfnamefont {T.}~\bibnamefont
  {Fukuhara}}, \bibinfo {author} {\bibfnamefont {S.}~\bibnamefont {Sugawa}},
  \bibinfo {author} {\bibfnamefont {Y.}~\bibnamefont {Takasu}}, \ and\ \bibinfo
  {author} {\bibfnamefont {Y.}~\bibnamefont {Takahashi}},\ }\bibfield  {title}
  {\enquote {\bibinfo {title} {{All-optical formation of quantum degenerate
  mixtures}},}\ }\href@noop {} {\bibfield  {journal} {\bibinfo  {journal}
  {Phys. Rev. A}\ }\textbf {\bibinfo {volume} {79}},\ \bibinfo {pages} {021601}
  (\bibinfo {year} {2009})}\BibitemShut {NoStop}%
\bibitem [{\citenamefont {Yamazaki}\ \emph {et~al.}(2010)\citenamefont
  {Yamazaki}, \citenamefont {Taie}, \citenamefont {Sugawa},\ and\ \citenamefont
  {Takahashi}}]{Yamazaki:2010en}%
  \BibitemOpen
  \bibfield  {author} {\bibinfo {author} {\bibfnamefont {R.}~\bibnamefont
  {Yamazaki}}, \bibinfo {author} {\bibfnamefont {S.}~\bibnamefont {Taie}},
  \bibinfo {author} {\bibfnamefont {S.}~\bibnamefont {Sugawa}}, \ and\ \bibinfo
  {author} {\bibfnamefont {Y.}~\bibnamefont {Takahashi}},\ }\bibfield  {title}
  {\enquote {\bibinfo {title} {{Submicron Spatial Modulation of an Interatomic
  Interaction in a Bose-Einstein Condensate}},}\ }\href@noop {} {\bibfield
  {journal} {\bibinfo  {journal} {Phys. Rev. Lett.}\ }\textbf {\bibinfo
  {volume} {105}},\ \bibinfo {pages} {050405} (\bibinfo {year}
  {2010})}\BibitemShut {NoStop}%
\bibitem [{\citenamefont {Blatt}\ \emph {et~al.}(2011)\citenamefont {Blatt},
  \citenamefont {Nicholson}, \citenamefont {Bloom}, \citenamefont {Williams},
  \citenamefont {Thomsen}, \citenamefont {Julienne},\ and\ \citenamefont
  {Ye}}]{Blatt:2011dr}%
  \BibitemOpen
  \bibfield  {author} {\bibinfo {author} {\bibfnamefont {S.}~\bibnamefont
  {Blatt}}, \bibinfo {author} {\bibfnamefont {T.~L.}\ \bibnamefont
  {Nicholson}}, \bibinfo {author} {\bibfnamefont {B.~J.}\ \bibnamefont
  {Bloom}}, \bibinfo {author} {\bibfnamefont {J.~R.}\ \bibnamefont {Williams}},
  \bibinfo {author} {\bibfnamefont {J.~W.}\ \bibnamefont {Thomsen}}, \bibinfo
  {author} {\bibfnamefont {P.~S.}\ \bibnamefont {Julienne}}, \ and\ \bibinfo
  {author} {\bibfnamefont {J.}~\bibnamefont {Ye}},\ }\bibfield  {title}
  {\enquote {\bibinfo {title} {{Measurement of Optical Feshbach Resonances in
  an Ideal Gas}},}\ }\href@noop {} {\bibfield  {journal} {\bibinfo  {journal}
  {Phys. Rev. Lett.}\ }\textbf {\bibinfo {volume} {107}},\ \bibinfo {pages}
  {073202} (\bibinfo {year} {2011})}\BibitemShut {NoStop}%
\bibitem [{\citenamefont {Wu}\ and\ \citenamefont {Thomas}(2012)}]{Wu:2012jp}%
  \BibitemOpen
  \bibfield  {author} {\bibinfo {author} {\bibfnamefont {H.}~\bibnamefont
  {Wu}}\ and\ \bibinfo {author} {\bibfnamefont {J.~E.}\ \bibnamefont
  {Thomas}},\ }\bibfield  {title} {\enquote {\bibinfo {title} {{Optical Control
  of Feshbach Resonances in Fermi Gases Using Molecular Dark States}},}\
  }\href@noop {} {\bibfield  {journal} {\bibinfo  {journal} {Phys. Rev. Lett.}\
  }\textbf {\bibinfo {volume} {108}},\ \bibinfo {pages} {010401} (\bibinfo
  {year} {2012})}\BibitemShut {NoStop}%
\bibitem [{\citenamefont {Jo}\ \emph {et~al.}(2012)\citenamefont {Jo},
  \citenamefont {Guzman}, \citenamefont {Thomas}, \citenamefont {Hosur},
  \citenamefont {Vishwanath},\ and\ \citenamefont {Stamper-Kurn}}]{Jo:2012bra}%
  \BibitemOpen
  \bibfield  {author} {\bibinfo {author} {\bibfnamefont {G.-B.}\ \bibnamefont
  {Jo}}, \bibinfo {author} {\bibfnamefont {J.}~\bibnamefont {Guzman}}, \bibinfo
  {author} {\bibfnamefont {C.~K.}\ \bibnamefont {Thomas}}, \bibinfo {author}
  {\bibfnamefont {P.}~\bibnamefont {Hosur}}, \bibinfo {author} {\bibfnamefont
  {A.}~\bibnamefont {Vishwanath}}, \ and\ \bibinfo {author} {\bibfnamefont
  {D.~M.}\ \bibnamefont {Stamper-Kurn}},\ }\bibfield  {title} {\enquote
  {\bibinfo {title} {{Ultracold Atoms in a Tunable Optical Kagome Lattice}},}\
  }\href@noop {} {\bibfield  {journal} {\bibinfo  {journal} {Phys. Rev. Lett.}\
  }\textbf {\bibinfo {volume} {108}},\ \bibinfo {pages} {045305} (\bibinfo
  {year} {2012})}\BibitemShut {NoStop}%
\bibitem [{\citenamefont {Tarruell}\ \emph {et~al.}(2012)\citenamefont
  {Tarruell}, \citenamefont {Greif}, \citenamefont {Uehlinger}, \citenamefont
  {Jotzu},\ and\ \citenamefont {Esslinger}}]{Tarruell:2012db}%
  \BibitemOpen
  \bibfield  {author} {\bibinfo {author} {\bibfnamefont {L.}~\bibnamefont
  {Tarruell}}, \bibinfo {author} {\bibfnamefont {D.}~\bibnamefont {Greif}},
  \bibinfo {author} {\bibfnamefont {T.}~\bibnamefont {Uehlinger}}, \bibinfo
  {author} {\bibfnamefont {G.}~\bibnamefont {Jotzu}}, \ and\ \bibinfo {author}
  {\bibfnamefont {T.}~\bibnamefont {Esslinger}},\ }\bibfield  {title} {\enquote
  {\bibinfo {title} {{Creating, moving and merging Dirac points with a Fermi
  gas in a tunable honeycomb lattice}},}\ }\href@noop {} {\bibfield  {journal}
  {\bibinfo  {journal} {Nature}\ }\textbf {\bibinfo {volume} {483}},\ \bibinfo
  {pages} {302--305} (\bibinfo {year} {2012})}\BibitemShut {NoStop}%
\bibitem [{\citenamefont {Wu}\ \emph {et~al.}(2007)\citenamefont {Wu},
  \citenamefont {Bergman}, \citenamefont {Balents},\ and\ \citenamefont
  {Das~Sarma}}]{Wu:2007iz}%
  \BibitemOpen
  \bibfield  {author} {\bibinfo {author} {\bibfnamefont {C.}~\bibnamefont
  {Wu}}, \bibinfo {author} {\bibfnamefont {D.}~\bibnamefont {Bergman}},
  \bibinfo {author} {\bibfnamefont {L.}~\bibnamefont {Balents}}, \ and\
  \bibinfo {author} {\bibfnamefont {S.}~\bibnamefont {Das~Sarma}},\ }\bibfield
  {title} {\enquote {\bibinfo {title} {Flat bands and wigner crystallization in
  the honeycomb optical lattice},}\ }\href {\doibase
  10.1103/PhysRevLett.99.070401} {\bibfield  {journal} {\bibinfo  {journal}
  {Phys. Rev. Lett.}\ }\textbf {\bibinfo {volume} {99}},\ \bibinfo {pages}
  {070401} (\bibinfo {year} {2007})}\BibitemShut {NoStop}%
\bibitem [{\citenamefont {Cooper}\ and\ \citenamefont
  {Dalibard}(2013)}]{Cooper:2013jg}%
  \BibitemOpen
  \bibfield  {author} {\bibinfo {author} {\bibfnamefont {N.~R.}\ \bibnamefont
  {Cooper}}\ and\ \bibinfo {author} {\bibfnamefont {J.}~\bibnamefont
  {Dalibard}},\ }\bibfield  {title} {\enquote {\bibinfo {title} {{Reaching
  Fractional Quantum Hall States with Optical Flux Lattices}},}\ }\href@noop {}
  {\bibfield  {journal} {\bibinfo  {journal} {Phys. Rev. Lett.}\ }\textbf
  {\bibinfo {volume} {110}},\ \bibinfo {pages} {185301} (\bibinfo {year}
  {2013})}\BibitemShut {NoStop}%
\bibitem [{\citenamefont {Paananen}\ and\ \citenamefont
  {Dahm}(2015)}]{Paananen:2015ie}%
  \BibitemOpen
  \bibfield  {author} {\bibinfo {author} {\bibfnamefont {T.}~\bibnamefont
  {Paananen}}\ and\ \bibinfo {author} {\bibfnamefont {T.}~\bibnamefont
  {Dahm}},\ }\bibfield  {title} {\enquote {\bibinfo {title} {{Topological flat
  bands in optical checkerboardlike lattices}},}\ }\href@noop {} {\bibfield
  {journal} {\bibinfo  {journal} {Phys. Rev. A}\ }\textbf {\bibinfo {volume}
  {91}},\ \bibinfo {pages} {033604} (\bibinfo {year} {2015})}\BibitemShut
  {NoStop}%
\bibitem [{\citenamefont {Chern}\ \emph {et~al.}(2014)\citenamefont {Chern},
  \citenamefont {Chien},\ and\ \citenamefont {Di~Ventra}}]{Chern:2014wg}%
  \BibitemOpen
  \bibfield  {author} {\bibinfo {author} {\bibfnamefont {G.-W.}\ \bibnamefont
  {Chern}}, \bibinfo {author} {\bibfnamefont {C.-C.}\ \bibnamefont {Chien}}, \
  and\ \bibinfo {author} {\bibfnamefont {M.}~\bibnamefont {Di~Ventra}},\
  }\bibfield  {title} {\enquote {\bibinfo {title} {{Dynamically generated
  flat-band phases in optical kagome lattices}},}\ }\href@noop {} {\bibfield
  {journal} {\bibinfo  {journal} {Phys. Rev. A}\ }\textbf {\bibinfo {volume}
  {90}},\ \bibinfo {pages} {013609} (\bibinfo {year} {2014})}\BibitemShut
  {NoStop}%
\bibitem [{\citenamefont {Chien}\ \emph {et~al.}(2012)\citenamefont {Chien},
  \citenamefont {Zwolak},\ and\ \citenamefont {Di~Ventra}}]{Chien:2012ft}%
  \BibitemOpen
  \bibfield  {author} {\bibinfo {author} {\bibfnamefont {C.-C.}\ \bibnamefont
  {Chien}}, \bibinfo {author} {\bibfnamefont {M.}~\bibnamefont {Zwolak}}, \
  and\ \bibinfo {author} {\bibfnamefont {M.}~\bibnamefont {Di~Ventra}},\
  }\bibfield  {title} {\enquote {\bibinfo {title} {{Bosonic and fermionic
  transport phenomena of ultracold atoms in one-dimensional optical
  lattices}},}\ }\href@noop {} {\bibfield  {journal} {\bibinfo  {journal}
  {Phys. Rev. A}\ }\textbf {\bibinfo {volume} {85}},\ \bibinfo {pages} {041601}
  (\bibinfo {year} {2012})}\BibitemShut {NoStop}%
\bibitem [{\citenamefont {Seaman}\ \emph {et~al.}(2007)\citenamefont {Seaman},
  \citenamefont {Kr{\"a}mer}, \citenamefont {Anderson},\ and\ \citenamefont
  {Holland}}]{Seaman:2007kx}%
  \BibitemOpen
  \bibfield  {author} {\bibinfo {author} {\bibfnamefont {B.~T.}\ \bibnamefont
  {Seaman}}, \bibinfo {author} {\bibfnamefont {M.}~\bibnamefont {Kr{\"a}mer}},
  \bibinfo {author} {\bibfnamefont {D.~Z.}\ \bibnamefont {Anderson}}, \ and\
  \bibinfo {author} {\bibfnamefont {M.~J.}\ \bibnamefont {Holland}},\
  }\bibfield  {title} {\enquote {\bibinfo {title} {{Atomtronics: Ultracold-atom
  analogs of electronic devices}},}\ }\href@noop {} {\bibfield  {journal}
  {\bibinfo  {journal} {Phys. Rev. A}\ }\textbf {\bibinfo {volume} {75}},\
  \bibinfo {pages} {023615} (\bibinfo {year} {2007})}\BibitemShut {NoStop}%
\bibitem [{\citenamefont {Pepino}\ \emph {et~al.}(2009)\citenamefont {Pepino},
  \citenamefont {Cooper}, \citenamefont {Anderson},\ and\ \citenamefont
  {Holland}}]{Pepino:2009jb}%
  \BibitemOpen
  \bibfield  {author} {\bibinfo {author} {\bibfnamefont {R.~A.}\ \bibnamefont
  {Pepino}}, \bibinfo {author} {\bibfnamefont {J.}~\bibnamefont {Cooper}},
  \bibinfo {author} {\bibfnamefont {D.~Z.}\ \bibnamefont {Anderson}}, \ and\
  \bibinfo {author} {\bibfnamefont {M.~J.}\ \bibnamefont {Holland}},\
  }\bibfield  {title} {\enquote {\bibinfo {title} {{Atomtronic Circuits of
  Diodes and Transistors}},}\ }\href@noop {} {\bibfield  {journal} {\bibinfo
  {journal} {Phys. Rev. Lett.}\ }\textbf {\bibinfo {volume} {103}},\ \bibinfo
  {pages} {140405} (\bibinfo {year} {2009})}\BibitemShut {NoStop}%
\bibitem [{\citenamefont {Di~Ventra}(2010)}]{DiVentra:2010ks}%
  \BibitemOpen
  \bibfield  {author} {\bibinfo {author} {\bibfnamefont {M.}~\bibnamefont
  {Di~Ventra}},\ }\href@noop {} {\emph {\bibinfo {title} {{Electrical Transport
  in Nanoscale Systems}}}}\ (\bibinfo  {publisher} {Cambridge University
  Press},\ \bibinfo {address} {Cambridge},\ \bibinfo {year} {2010})\BibitemShut
  {NoStop}%
\bibitem [{\citenamefont {Mahan}(2000)}]{mahan2000many}%
  \BibitemOpen
  \bibfield  {author} {\bibinfo {author} {\bibfnamefont {G.~D.}\ \bibnamefont
  {Mahan}},\ }\href@noop {} {\emph {\bibinfo {title} {{Many-Particle
  Physics}}}},\ Physics of Solids and Liquids\ (\bibinfo  {publisher}
  {Springer},\ \bibinfo {year} {2000})\BibitemShut {NoStop}%
\bibitem [{\citenamefont {Press}\ \emph {et~al.}(2007)\citenamefont {Press},
  \citenamefont {Teukolsky}, \citenamefont {Vetterling},\ and\ \citenamefont
  {Flannery}}]{NRE}%
  \BibitemOpen
  \bibfield  {author} {\bibinfo {author} {\bibfnamefont {W.~H.}\ \bibnamefont
  {Press}}, \bibinfo {author} {\bibfnamefont {S.~A.}\ \bibnamefont
  {Teukolsky}}, \bibinfo {author} {\bibfnamefont {W.~T.}\ \bibnamefont
  {Vetterling}}, \ and\ \bibinfo {author} {\bibfnamefont {B.~P.}\ \bibnamefont
  {Flannery}},\ }\href@noop {} {\emph {\bibinfo {title} {{Numerical Recipes 3rd
  Edition: The Art of Scientific Computing}}}},\ \bibinfo {edition} {3rd}\ ed.\
  (\bibinfo  {publisher} {Cambridge University Press},\ \bibinfo {address} {New
  York, NY, USA},\ \bibinfo {year} {2007})\BibitemShut {NoStop}%
\bibitem [{\citenamefont {Vudragovi{\'c}}\ \emph {et~al.}(2012)\citenamefont
  {Vudragovi{\'c}}, \citenamefont {Vidanovi{\'c}}, \citenamefont {Bala{\v z}},
  \citenamefont {Muruganandam},\ and\ \citenamefont
  {Adhikari}}]{Vudragovic:2012jz}%
  \BibitemOpen
  \bibfield  {author} {\bibinfo {author} {\bibfnamefont {D.}~\bibnamefont
  {Vudragovi{\'c}}}, \bibinfo {author} {\bibfnamefont {I.}~\bibnamefont
  {Vidanovi{\'c}}}, \bibinfo {author} {\bibfnamefont {A.}~\bibnamefont {Bala{\v
  z}}}, \bibinfo {author} {\bibfnamefont {P.}~\bibnamefont {Muruganandam}}, \
  and\ \bibinfo {author} {\bibfnamefont {S.~K.}\ \bibnamefont {Adhikari}},\
  }\bibfield  {title} {\enquote {\bibinfo {title} {{C programs for solving the
  time-dependent Gross{\textendash}Pitaevskii equation in a fully anisotropic
  trap}},}\ }\href@noop {} {\bibfield  {journal} {\bibinfo  {journal} {Computer
  Physics Communications}\ }\textbf {\bibinfo {volume} {183}},\ \bibinfo
  {pages} {2021--2025} (\bibinfo {year} {2012})}\BibitemShut {NoStop}%
\bibitem [{\citenamefont {Bao}\ and\ \citenamefont {Du}(2006)}]{Bao:2006fg}%
  \BibitemOpen
  \bibfield  {author} {\bibinfo {author} {\bibfnamefont {W.}~\bibnamefont
  {Bao}}\ and\ \bibinfo {author} {\bibfnamefont {Q.}~\bibnamefont {Du}},\
  }\bibfield  {title} {\enquote {\bibinfo {title} {{Computing the Ground State
  Solution of Bose--Einstein Condensates by a Normalized Gradient Flow}},}\
  }\href@noop {} {\bibfield  {journal} {\bibinfo  {journal} {SIAM Journal on
  Scientific Computing}\ }\textbf {\bibinfo {volume} {25}},\ \bibinfo {pages}
  {1674--1697} (\bibinfo {year} {2006})}\BibitemShut {NoStop}%
\bibitem [{\citenamefont {Chien}\ \emph {et~al.}(2013)\citenamefont {Chien},
  \citenamefont {Gruss}, \citenamefont {Di~Ventra},\ and\ \citenamefont
  {Zwolak}}]{Chien:2013ef}%
  \BibitemOpen
  \bibfield  {author} {\bibinfo {author} {\bibfnamefont {C.-C.}\ \bibnamefont
  {Chien}}, \bibinfo {author} {\bibfnamefont {D.}~\bibnamefont {Gruss}},
  \bibinfo {author} {\bibfnamefont {M.}~\bibnamefont {Di~Ventra}}, \ and\
  \bibinfo {author} {\bibfnamefont {M.}~\bibnamefont {Zwolak}},\ }\bibfield
  {title} {\enquote {\bibinfo {title} {{Interaction-induced
  conducting{\textendash}non-conducting transition of ultra-cold atoms in
  one-dimensional optical lattices}},}\ }\href@noop {} {\bibfield  {journal}
  {\bibinfo  {journal} {New J. Phys.}\ }\textbf {\bibinfo {volume} {15}},\
  \bibinfo {pages} {063026} (\bibinfo {year} {2013})}\BibitemShut {NoStop}%
\bibitem [{\citenamefont {Gemelke}\ \emph {et~al.}(2009)\citenamefont
  {Gemelke}, \citenamefont {Zhang}, \citenamefont {Hung},\ and\ \citenamefont
  {Chin}}]{Gemelke:2009ja}%
  \BibitemOpen
  \bibfield  {author} {\bibinfo {author} {\bibfnamefont {N.}~\bibnamefont
  {Gemelke}}, \bibinfo {author} {\bibfnamefont {X.}~\bibnamefont {Zhang}},
  \bibinfo {author} {\bibfnamefont {C.-L.}\ \bibnamefont {Hung}}, \ and\
  \bibinfo {author} {\bibfnamefont {C.}~\bibnamefont {Chin}},\ }\bibfield
  {title} {\enquote {\bibinfo {title} {{In situ observation of incompressible
  Mott-insulating domains in ultracold atomic gases}},}\ }\href@noop {}
  {\bibfield  {journal} {\bibinfo  {journal} {Nature}\ }\textbf {\bibinfo
  {volume} {460}},\ \bibinfo {pages} {995--998} (\bibinfo {year}
  {2009})}\BibitemShut {NoStop}%
\bibitem [{\citenamefont {Lee}\ \emph {et~al.}(2007)\citenamefont {Lee},
  \citenamefont {Anderlini}, \citenamefont {Brown}, \citenamefont
  {Sebby-Strabley}, \citenamefont {Phillips},\ and\ \citenamefont
  {Porto}}]{Lee:2007ip}%
  \BibitemOpen
  \bibfield  {author} {\bibinfo {author} {\bibfnamefont {P.~J.}\ \bibnamefont
  {Lee}}, \bibinfo {author} {\bibfnamefont {M.}~\bibnamefont {Anderlini}},
  \bibinfo {author} {\bibfnamefont {B.~L.}\ \bibnamefont {Brown}}, \bibinfo
  {author} {\bibfnamefont {J.}~\bibnamefont {Sebby-Strabley}}, \bibinfo
  {author} {\bibfnamefont {W.~D.}\ \bibnamefont {Phillips}}, \ and\ \bibinfo
  {author} {\bibfnamefont {J.~V.}\ \bibnamefont {Porto}},\ }\bibfield  {title}
  {\enquote {\bibinfo {title} {{Sublattice Addressing and Spin-Dependent Motion
  of Atoms in a Double-Well Lattice}},}\ }\href@noop {} {\bibfield  {journal}
  {\bibinfo  {journal} {Phys. Rev. Lett.}\ }\textbf {\bibinfo {volume} {99}},\
  \bibinfo {pages} {020402} (\bibinfo {year} {2007})}\BibitemShut {NoStop}%
\bibitem [{\citenamefont {Stewart}\ \emph {et~al.}(2008)\citenamefont
  {Stewart}, \citenamefont {Gaebler},\ and\ \citenamefont
  {Jin}}]{Stewart:2008kt}%
  \BibitemOpen
  \bibfield  {author} {\bibinfo {author} {\bibfnamefont {J.~T.}\ \bibnamefont
  {Stewart}}, \bibinfo {author} {\bibfnamefont {J.~P.}\ \bibnamefont
  {Gaebler}}, \ and\ \bibinfo {author} {\bibfnamefont {D.~S.}\ \bibnamefont
  {Jin}},\ }\bibfield  {title} {\enquote {\bibinfo {title} {{Using
  photoemission spectroscopy to probe a strongly interacting Fermi gas}},}\
  }\href@noop {} {\bibfield  {journal} {\bibinfo  {journal} {Nature}\ }\textbf
  {\bibinfo {volume} {454}},\ \bibinfo {pages} {744--747} (\bibinfo {year}
  {2008})}\BibitemShut {NoStop}%
\bibitem [{\citenamefont {Cerimele}\ \emph {et~al.}(2000)\citenamefont
  {Cerimele}, \citenamefont {Chiofalo}, \citenamefont {Pistella}, \citenamefont
  {Succi},\ and\ \citenamefont {Tosi}}]{Cerimele:2000hs}%
  \BibitemOpen
  \bibfield  {author} {\bibinfo {author} {\bibfnamefont {M.~M.}\ \bibnamefont
  {Cerimele}}, \bibinfo {author} {\bibfnamefont {M.~L.}\ \bibnamefont
  {Chiofalo}}, \bibinfo {author} {\bibfnamefont {F.}~\bibnamefont {Pistella}},
  \bibinfo {author} {\bibfnamefont {S.}~\bibnamefont {Succi}}, \ and\ \bibinfo
  {author} {\bibfnamefont {M.~P.}\ \bibnamefont {Tosi}},\ }\bibfield  {title}
  {\enquote {\bibinfo {title} {{Numerical solution of the Gross-Pitaevskii
  equation using an explicit finite-difference scheme: An application to
  trapped Bose-Einstein condensates}},}\ }\href@noop {} {\bibfield  {journal}
  {\bibinfo  {journal} {Phys. Rev. E}\ }\textbf {\bibinfo {volume} {62}},\
  \bibinfo {pages} {1382--1389} (\bibinfo {year} {2000})}\BibitemShut {NoStop}%
\bibitem [{\citenamefont {Peotta}\ \emph {et~al.}(2014)\citenamefont {Peotta},
  \citenamefont {Chien},\ and\ \citenamefont {Di~Ventra}}]{Peotta:2014woa}%
  \BibitemOpen
  \bibfield  {author} {\bibinfo {author} {\bibfnamefont {S.}~\bibnamefont
  {Peotta}}, \bibinfo {author} {\bibfnamefont {C.-C.}\ \bibnamefont {Chien}}, \
  and\ \bibinfo {author} {\bibfnamefont {M.}~\bibnamefont {Di~Ventra}},\
  }\bibfield  {title} {\enquote {\bibinfo {title} {{Phase-induced transport in
  atomic gases: From superfluid to Mott insulator}},}\ }\href@noop {}
  {\bibfield  {journal} {\bibinfo  {journal} {Phys. Rev. A}\ }\textbf {\bibinfo
  {volume} {90}},\ \bibinfo {pages} {053615} (\bibinfo {year}
  {2014})}\BibitemShut {NoStop}%
\bibitem [{\citenamefont {Henderson}\ \emph {et~al.}(2009)\citenamefont
  {Henderson}, \citenamefont {Ryu}, \citenamefont {MacCormick},\ and\
  \citenamefont {Boshier}}]{Henderson:2009eo}%
  \BibitemOpen
  \bibfield  {author} {\bibinfo {author} {\bibfnamefont {K.}~\bibnamefont
  {Henderson}}, \bibinfo {author} {\bibfnamefont {C.}~\bibnamefont {Ryu}},
  \bibinfo {author} {\bibfnamefont {C.}~\bibnamefont {MacCormick}}, \ and\
  \bibinfo {author} {\bibfnamefont {M.~G.}\ \bibnamefont {Boshier}},\
  }\bibfield  {title} {\enquote {\bibinfo {title} {{Experimental demonstration
  of painting arbitrary and dynamic potentials for Bose{\textendash}Einstein
  condensates}},}\ }\href@noop {} {\bibfield  {journal} {\bibinfo  {journal}
  {New J. Phys.}\ }\textbf {\bibinfo {volume} {11}},\ \bibinfo {pages} {043030}
  (\bibinfo {year} {2009})}\BibitemShut {NoStop}%
\bibitem [{\citenamefont {Gericke}\ \emph {et~al.}(2008)\citenamefont
  {Gericke}, \citenamefont {W{\"u}rtz}, \citenamefont {Reitz}, \citenamefont
  {Langen},\ and\ \citenamefont {Ott}}]{Gericke:2008jw}%
  \BibitemOpen
  \bibfield  {author} {\bibinfo {author} {\bibfnamefont {T.}~\bibnamefont
  {Gericke}}, \bibinfo {author} {\bibfnamefont {P.}~\bibnamefont {W{\"u}rtz}},
  \bibinfo {author} {\bibfnamefont {D.}~\bibnamefont {Reitz}}, \bibinfo
  {author} {\bibfnamefont {T.}~\bibnamefont {Langen}}, \ and\ \bibinfo {author}
  {\bibfnamefont {H.}~\bibnamefont {Ott}},\ }\bibfield  {title} {\enquote
  {\bibinfo {title} {{High-resolution scanning electron microscopy of an
  ultracold quantum gas}},}\ }\href@noop {} {\bibfield  {journal} {\bibinfo
  {journal} {Nat Phys}\ }\textbf {\bibinfo {volume} {4}},\ \bibinfo {pages}
  {949--953} (\bibinfo {year} {2008})}\BibitemShut {NoStop}%
\bibitem [{\citenamefont {Patil}\ \emph {et~al.}(2014)\citenamefont {Patil},
  \citenamefont {Chakram}, \citenamefont {Aycock},\ and\ \citenamefont
  {Vengalattore}}]{Patil:2014jv}%
  \BibitemOpen
  \bibfield  {author} {\bibinfo {author} {\bibfnamefont {Y.~S.}\ \bibnamefont
  {Patil}}, \bibinfo {author} {\bibfnamefont {S.}~\bibnamefont {Chakram}},
  \bibinfo {author} {\bibfnamefont {L.~M.}\ \bibnamefont {Aycock}}, \ and\
  \bibinfo {author} {\bibfnamefont {M.}~\bibnamefont {Vengalattore}},\
  }\bibfield  {title} {\enquote {\bibinfo {title} {{Nondestructive imaging of
  an ultracold lattice gas}},}\ }\href@noop {} {\bibfield  {journal} {\bibinfo
  {journal} {Phys. Rev. A}\ }\textbf {\bibinfo {volume} {90}},\ \bibinfo
  {pages} {033422} (\bibinfo {year} {2014})}\BibitemShut {NoStop}%
\bibitem [{\citenamefont {Wang}\ \emph {et~al.}(2014)\citenamefont {Wang},
  \citenamefont {Deng},\ and\ \citenamefont {Duan}}]{Wang:2014ke}%
  \BibitemOpen
  \bibfield  {author} {\bibinfo {author} {\bibfnamefont {S.~T.}\ \bibnamefont
  {Wang}}, \bibinfo {author} {\bibfnamefont {D.~L.}\ \bibnamefont {Deng}}, \
  and\ \bibinfo {author} {\bibfnamefont {L.~M.}\ \bibnamefont {Duan}},\
  }\bibfield  {title} {\enquote {\bibinfo {title} {{Probe of Three-Dimensional
  Chiral Topological Insulators in an Optical Lattice}},}\ }\href@noop {}
  {\bibfield  {journal} {\bibinfo  {journal} {Phys. Rev. Lett.}\ }\textbf
  {\bibinfo {volume} {113}},\ \bibinfo {pages} {033002} (\bibinfo {year}
  {2014})}\BibitemShut {NoStop}%
\bibitem [{\citenamefont {Hart}\ \emph {et~al.}(2015)\citenamefont {Hart},
  \citenamefont {Duarte}, \citenamefont {Yang}, \citenamefont {Liu},
  \citenamefont {Paiva}, \citenamefont {Khatami}, \citenamefont {Scalettar},
  \citenamefont {Trivedi}, \citenamefont {Huse},\ and\ \citenamefont
  {Hulet}}]{Hart:2015ex}%
  \BibitemOpen
  \bibfield  {author} {\bibinfo {author} {\bibfnamefont {R.~A.}\ \bibnamefont
  {Hart}}, \bibinfo {author} {\bibfnamefont {P.~M.}\ \bibnamefont {Duarte}},
  \bibinfo {author} {\bibfnamefont {T.-L.}\ \bibnamefont {Yang}}, \bibinfo
  {author} {\bibfnamefont {X.}~\bibnamefont {Liu}}, \bibinfo {author}
  {\bibfnamefont {T.}~\bibnamefont {Paiva}}, \bibinfo {author} {\bibfnamefont
  {E.}~\bibnamefont {Khatami}}, \bibinfo {author} {\bibfnamefont {R.~T.}\
  \bibnamefont {Scalettar}}, \bibinfo {author} {\bibfnamefont {N.}~\bibnamefont
  {Trivedi}}, \bibinfo {author} {\bibfnamefont {D.~A.}\ \bibnamefont {Huse}}, \
  and\ \bibinfo {author} {\bibfnamefont {R.~G.}\ \bibnamefont {Hulet}},\
  }\bibfield  {title} {\enquote {\bibinfo {title} {{Observation of
  antiferromagnetic correlations in the Hubbard model with ultracold atoms}},}\
  }\href@noop {} {\bibfield  {journal} {\bibinfo  {journal} {Nature}\ }\textbf
  {\bibinfo {volume} {519}},\ \bibinfo {pages} {211--214} (\bibinfo {year}
  {2015})}\BibitemShut {NoStop}%
\bibitem [{\citenamefont {Duarte}\ \emph {et~al.}(2015)\citenamefont {Duarte},
  \citenamefont {Hart}, \citenamefont {Yang}, \citenamefont {Liu},
  \citenamefont {Paiva}, \citenamefont {Khatami}, \citenamefont {Scalettar},
  \citenamefont {Trivedi},\ and\ \citenamefont {Hulet}}]{Duarte:2015cn}%
  \BibitemOpen
  \bibfield  {author} {\bibinfo {author} {\bibfnamefont {P.~M.}\ \bibnamefont
  {Duarte}}, \bibinfo {author} {\bibfnamefont {R.~A.}\ \bibnamefont {Hart}},
  \bibinfo {author} {\bibfnamefont {T.-L.}\ \bibnamefont {Yang}}, \bibinfo
  {author} {\bibfnamefont {X.}~\bibnamefont {Liu}}, \bibinfo {author}
  {\bibfnamefont {T.}~\bibnamefont {Paiva}}, \bibinfo {author} {\bibfnamefont
  {E.}~\bibnamefont {Khatami}}, \bibinfo {author} {\bibfnamefont {R.~T.}\
  \bibnamefont {Scalettar}}, \bibinfo {author} {\bibfnamefont {N.}~\bibnamefont
  {Trivedi}}, \ and\ \bibinfo {author} {\bibfnamefont {R.~G.}\ \bibnamefont
  {Hulet}},\ }\bibfield  {title} {\enquote {\bibinfo {title} {{Compressibility
  of a Fermionic Mott Insulator of Ultracold Atoms}},}\ }\href@noop {}
  {\bibfield  {journal} {\bibinfo  {journal} {Phys. Rev. Lett.}\ }\textbf
  {\bibinfo {volume} {114}},\ \bibinfo {pages} {070403} (\bibinfo {year}
  {2015})}\BibitemShut {NoStop}%
\bibitem [{\citenamefont {Lester}\ \emph {et~al.}(2015)\citenamefont {Lester},
  \citenamefont {Luick}, \citenamefont {Kaufman}, \citenamefont {Reynolds},\
  and\ \citenamefont {Regal}}]{Lester:2015if}%
  \BibitemOpen
  \bibfield  {author} {\bibinfo {author} {\bibfnamefont {B.~J.}\ \bibnamefont
  {Lester}}, \bibinfo {author} {\bibfnamefont {N.}~\bibnamefont {Luick}},
  \bibinfo {author} {\bibfnamefont {A.~M.}\ \bibnamefont {Kaufman}}, \bibinfo
  {author} {\bibfnamefont {C.~M.}\ \bibnamefont {Reynolds}}, \ and\ \bibinfo
  {author} {\bibfnamefont {C.~A.}\ \bibnamefont {Regal}},\ }\bibfield  {title}
  {\enquote {\bibinfo {title} {Rapid production of uniformly filled arrays of
  neutral atoms},}\ }\href {\doibase 10.1103/PhysRevLett.115.073003} {\bibfield
   {journal} {\bibinfo  {journal} {Phys. Rev. Lett.}\ }\textbf {\bibinfo
  {volume} {115}},\ \bibinfo {pages} {073003} (\bibinfo {year}
  {2015})}\BibitemShut {NoStop}%
\bibitem [{\citenamefont {Beugnon}\ \emph {et~al.}(2007)\citenamefont
  {Beugnon}, \citenamefont {Tuchendler}, \citenamefont {Marion}, \citenamefont
  {Ga{\"e}tan}, \citenamefont {Miroshnychenko}, \citenamefont {Sortais},
  \citenamefont {Lance}, \citenamefont {Jones}, \citenamefont {Messin},
  \citenamefont {Browaeys},\ and\ \citenamefont {Grangier}}]{Beugnon:2007eg}%
  \BibitemOpen
  \bibfield  {author} {\bibinfo {author} {\bibfnamefont {J.}~\bibnamefont
  {Beugnon}}, \bibinfo {author} {\bibfnamefont {C.}~\bibnamefont {Tuchendler}},
  \bibinfo {author} {\bibfnamefont {H.}~\bibnamefont {Marion}}, \bibinfo
  {author} {\bibfnamefont {A.}~\bibnamefont {Ga{\"e}tan}}, \bibinfo {author}
  {\bibfnamefont {Y.}~\bibnamefont {Miroshnychenko}}, \bibinfo {author}
  {\bibfnamefont {Y.~R.~P.}\ \bibnamefont {Sortais}}, \bibinfo {author}
  {\bibfnamefont {A.~M.}\ \bibnamefont {Lance}}, \bibinfo {author}
  {\bibfnamefont {M.~P.~A.}\ \bibnamefont {Jones}}, \bibinfo {author}
  {\bibfnamefont {G.}~\bibnamefont {Messin}}, \bibinfo {author} {\bibfnamefont
  {A.}~\bibnamefont {Browaeys}}, \ and\ \bibinfo {author} {\bibfnamefont
  {P.}~\bibnamefont {Grangier}},\ }\bibfield  {title} {\enquote {\bibinfo
  {title} {Two-dimensional transport and transfer of a single atomic qubit in
  optical tweezers},}\ }\href@noop {} {\bibfield  {journal} {\bibinfo
  {journal} {Nat. Phys.}\ }\textbf {\bibinfo {volume} {3}},\ \bibinfo {pages}
  {696--699} (\bibinfo {year} {2007})}\BibitemShut {NoStop}%
\bibitem [{\citenamefont {Weiss}\ \emph {et~al.}(2004)\citenamefont {Weiss},
  \citenamefont {Vala}, \citenamefont {Thapliyal}, \citenamefont {Myrgren},
  \citenamefont {Vazirani},\ and\ \citenamefont {Whaley}}]{Weiss:2004fk}%
  \BibitemOpen
  \bibfield  {author} {\bibinfo {author} {\bibfnamefont {D.~S.}\ \bibnamefont
  {Weiss}}, \bibinfo {author} {\bibfnamefont {J.}~\bibnamefont {Vala}},
  \bibinfo {author} {\bibfnamefont {A.~V.}\ \bibnamefont {Thapliyal}}, \bibinfo
  {author} {\bibfnamefont {S.}~\bibnamefont {Myrgren}}, \bibinfo {author}
  {\bibfnamefont {U.}~\bibnamefont {Vazirani}}, \ and\ \bibinfo {author}
  {\bibfnamefont {K.~B.}\ \bibnamefont {Whaley}},\ }\bibfield  {title}
  {\enquote {\bibinfo {title} {{Another way to approach zero entropy for a
  finite system of atoms}},}\ }\href@noop {} {\bibfield  {journal} {\bibinfo
  {journal} {Phys. Rev. A}\ }\textbf {\bibinfo {volume} {70}},\ \bibinfo
  {pages} {040302} (\bibinfo {year} {2004})}\BibitemShut {NoStop}%
\bibitem [{\citenamefont {Horowitz}\ and\ \citenamefont
  {Hill}(1989)}]{HorowitzBook}%
  \BibitemOpen
  \bibfield  {author} {\bibinfo {author} {\bibfnamefont {P.}~\bibnamefont
  {Horowitz}}\ and\ \bibinfo {author} {\bibfnamefont {W.}~\bibnamefont
  {Hill}},\ }\href@noop {} {\emph {\bibinfo {title} {The art of
  electronics}}},\ \bibinfo {edition} {2nd}\ ed.\ (\bibinfo  {publisher}
  {Cambridge University Press},\ \bibinfo {address} {Cambridge},\ \bibinfo
  {year} {1989})\BibitemShut {NoStop}%
\bibitem [{\citenamefont {Rushton}\ \emph {et~al.}(2014)\citenamefont
  {Rushton}, \citenamefont {Aldous},\ and\ \citenamefont
  {Himsworth}}]{Rushton14}%
  \BibitemOpen
  \bibfield  {author} {\bibinfo {author} {\bibfnamefont {J.}~\bibnamefont
  {Rushton}}, \bibinfo {author} {\bibfnamefont {M.}~\bibnamefont {Aldous}}, \
  and\ \bibinfo {author} {\bibfnamefont {M.}~\bibnamefont {Himsworth}},\
  }\bibfield  {title} {\enquote {\bibinfo {title} {The feasibility of a fully
  miniaturized magneto-optical trap for portable ultracold quantum
  technology},}\ }\href
  {http://scitation.aip.org/content/aip/journal/rsi/85/12/10.1063/1.4904066}
  {\bibfield  {journal} {\bibinfo  {journal} {Rev. Sci. Instrum.}\ }\textbf
  {\bibinfo {volume} {85}},\ \bibinfo {pages} {121501} (\bibinfo {year}
  {2014})}\BibitemShut {NoStop}%
\bibitem [{\citenamefont {Kester}\ and\ \citenamefont
  {Analog~Devices}(2005)}]{kester2005data}%
  \BibitemOpen
  \bibfield  {author} {\bibinfo {author} {\bibfnamefont {W.}~\bibnamefont
  {Kester}}\ and\ \bibinfo {author} {\bibfnamefont {i.}~\bibnamefont
  {Analog~Devices}},\ }\href {https://books.google.com/books?id=0aeBS6SgtR4C}
  {\emph {\bibinfo {title} {Data Conversion Handbook}}},\ Analog Devices
  series\ (\bibinfo  {publisher} {Elsevier},\ \bibinfo {year}
  {2005})\BibitemShut {NoStop}%
\bibitem [{\citenamefont {Moore}\ \emph {et~al.}(2015)\citenamefont {Moore},
  \citenamefont {Lee}, \citenamefont {Findlay}, \citenamefont {Torralbo-Campo},
  \citenamefont {Bruce},\ and\ \citenamefont {Cassettari}}]{Moore:2015kd}%
  \BibitemOpen
  \bibfield  {author} {\bibinfo {author} {\bibfnamefont {R.~W.~G.}\
  \bibnamefont {Moore}}, \bibinfo {author} {\bibfnamefont {L.~A.}\ \bibnamefont
  {Lee}}, \bibinfo {author} {\bibfnamefont {E.~A.}\ \bibnamefont {Findlay}},
  \bibinfo {author} {\bibfnamefont {L.}~\bibnamefont {Torralbo-Campo}},
  \bibinfo {author} {\bibfnamefont {G.~D.}\ \bibnamefont {Bruce}}, \ and\
  \bibinfo {author} {\bibfnamefont {D.}~\bibnamefont {Cassettari}},\ }\bibfield
   {title} {\enquote {\bibinfo {title} {{Measurement of vacuum pressure with a
  magneto-optical trap: A pressure-rise method}},}\ }\href@noop {} {\bibfield
  {journal} {\bibinfo  {journal} {Review of Scientific Instruments}\ }\textbf
  {\bibinfo {volume} {86}},\ \bibinfo {pages} {093108} (\bibinfo {year}
  {2015})}\BibitemShut {NoStop}%
\bibitem [{\citenamefont {Zhang}\ and\ \citenamefont
  {Jo}(2015)}]{Zhang:2015vg}%
  \BibitemOpen
  \bibfield  {author} {\bibinfo {author} {\bibfnamefont {T.}~\bibnamefont
  {Zhang}}\ and\ \bibinfo {author} {\bibfnamefont {G.~B.}\ \bibnamefont {Jo}},\
  }\bibfield  {title} {\enquote {\bibinfo {title} {One-dimensional sawtooth and
  zigzag lattices for ultracold atoms},}\ }\href@noop {} {\bibfield  {journal}
  {\bibinfo  {journal} {Sci. Rep.}\ }\textbf {\bibinfo {volume} {5}},\ \bibinfo
  {pages} {16044} (\bibinfo {year} {2015})}\BibitemShut {NoStop}%
\bibitem [{\citenamefont {Flach}\ \emph {et~al.}(2014)\citenamefont {Flach},
  \citenamefont {Leykam}, \citenamefont {Bodyfelt}, \citenamefont {Matthies},\
  and\ \citenamefont {Desyatnikov}}]{Flach:2014cm}%
  \BibitemOpen
  \bibfield  {author} {\bibinfo {author} {\bibfnamefont {S.}~\bibnamefont
  {Flach}}, \bibinfo {author} {\bibfnamefont {D.}~\bibnamefont {Leykam}},
  \bibinfo {author} {\bibfnamefont {J.~D.}\ \bibnamefont {Bodyfelt}}, \bibinfo
  {author} {\bibfnamefont {P.}~\bibnamefont {Matthies}}, \ and\ \bibinfo
  {author} {\bibfnamefont {A.~S.}\ \bibnamefont {Desyatnikov}},\ }\bibfield
  {title} {\enquote {\bibinfo {title} {{Detangling flat bands into Fano
  lattices}},}\ }\href@noop {} {\bibfield  {journal} {\bibinfo  {journal}
  {EPL}\ }\textbf {\bibinfo {volume} {105}},\ \bibinfo {pages} {30001}
  (\bibinfo {year} {2014})}\BibitemShut {NoStop}%
\bibitem [{\citenamefont {Metcalf}\ \emph {et~al.}(2015)\citenamefont
  {Metcalf}, \citenamefont {Chern}, \citenamefont {Di~Ventra},\ and\
  \citenamefont {Chien}}]{Metcalf:2015wj}%
  \BibitemOpen
  \bibfield  {author} {\bibinfo {author} {\bibfnamefont {M.}~\bibnamefont
  {Metcalf}}, \bibinfo {author} {\bibfnamefont {G.-W.}\ \bibnamefont {Chern}},
  \bibinfo {author} {\bibfnamefont {M.}~\bibnamefont {Di~Ventra}}, \ and\
  \bibinfo {author} {\bibfnamefont {C.-C.}\ \bibnamefont {Chien}},\ }\bibfield
  {title} {\enquote {\bibinfo {title} {Matter-wave propagation in optical
  lattices: geometrical and flat-band effects},}\ }\href@noop {} {\bibfield
  {journal} {\bibinfo  {journal} {arXiv: 1502.04975}\ } (\bibinfo {year}
  {2015})}\BibitemShut {NoStop}%
\bibitem [{\citenamefont {Mukherjee}\ and\ \citenamefont
  {Thomson}(2015)}]{Mukherjee:2015uf}%
  \BibitemOpen
  \bibfield  {author} {\bibinfo {author} {\bibfnamefont {S.}~\bibnamefont
  {Mukherjee}}\ and\ \bibinfo {author} {\bibfnamefont {R.~R.}\ \bibnamefont
  {Thomson}},\ }\bibfield  {title} {\enquote {\bibinfo {title} {Observation of
  localized flat-band modes in a one-dimensional photonic rhombic lattice},}\
  }\href@noop {} {\bibfield  {journal} {\bibinfo  {journal} {Opt. Lett.}\
  }\textbf {\bibinfo {volume} {40}},\ \bibinfo {pages} {5443} (\bibinfo {year}
  {2015})}\BibitemShut {NoStop}%
\bibitem [{\citenamefont {Taie}\ \emph {et~al.}(2015)\citenamefont {Taie},
  \citenamefont {Ozawa}, \citenamefont {Ichinose}, \citenamefont {Nishio},
  \citenamefont {Nakajima},\ and\ \citenamefont {Takahashi}}]{Taie15}%
  \BibitemOpen
  \bibfield  {author} {\bibinfo {author} {\bibfnamefont {S.}~\bibnamefont
  {Taie}}, \bibinfo {author} {\bibfnamefont {H.}~\bibnamefont {Ozawa}},
  \bibinfo {author} {\bibfnamefont {T.}~\bibnamefont {Ichinose}}, \bibinfo
  {author} {\bibfnamefont {T.}~\bibnamefont {Nishio}}, \bibinfo {author}
  {\bibfnamefont {S.}~\bibnamefont {Nakajima}}, \ and\ \bibinfo {author}
  {\bibfnamefont {Y.}~\bibnamefont {Takahashi}},\ }\bibfield  {title} {\enquote
  {\bibinfo {title} {Matter-wave localization and delocalization of ultracold
  bosons in an optical lieb lattice},}\ }\href@noop {} {\bibfield  {journal}
  {\bibinfo  {journal} {arXiv:1506.00587}\ } (\bibinfo {year}
  {2015})}\BibitemShut {NoStop}%
\bibitem [{\citenamefont {Vicencio}\ \emph {et~al.}(2015)\citenamefont
  {Vicencio}, \citenamefont {Cantillano}, \citenamefont {Morales-Inostroza},
  \citenamefont {Real}, \citenamefont {Mejia-Cortes}, \citenamefont {Weimann},
  \citenamefont {Szameit},\ and\ \citenamefont {Molina}}]{Vicencio:2015ks}%
  \BibitemOpen
  \bibfield  {author} {\bibinfo {author} {\bibfnamefont {R.~A.}\ \bibnamefont
  {Vicencio}}, \bibinfo {author} {\bibfnamefont {C.}~\bibnamefont
  {Cantillano}}, \bibinfo {author} {\bibfnamefont {L.}~\bibnamefont
  {Morales-Inostroza}}, \bibinfo {author} {\bibfnamefont {B.}~\bibnamefont
  {Real}}, \bibinfo {author} {\bibfnamefont {C.}~\bibnamefont {Mejia-Cortes}},
  \bibinfo {author} {\bibfnamefont {S.}~\bibnamefont {Weimann}}, \bibinfo
  {author} {\bibfnamefont {A.}~\bibnamefont {Szameit}}, \ and\ \bibinfo
  {author} {\bibfnamefont {M.~I.}\ \bibnamefont {Molina}},\ }\bibfield  {title}
  {\enquote {\bibinfo {title} {{Observation of Localized States in Lieb
  Photonic Lattices}},}\ }\href@noop {} {\bibfield  {journal} {\bibinfo
  {journal} {Phys. Rev. Lett.}\ }\textbf {\bibinfo {volume} {114}},\ \bibinfo
  {pages} {245503} (\bibinfo {year} {2015})}\BibitemShut {NoStop}%
\bibitem [{\citenamefont {Mukherjee}\ \emph {et~al.}(2015)\citenamefont
  {Mukherjee}, \citenamefont {Spracklen}, \citenamefont {Choudhury},
  \citenamefont {Goldman}, \citenamefont {{\"O}hberg}, \citenamefont
  {Andersson},\ and\ \citenamefont {Thomson}}]{Mukherjee:2015iu}%
  \BibitemOpen
  \bibfield  {author} {\bibinfo {author} {\bibfnamefont {S.}~\bibnamefont
  {Mukherjee}}, \bibinfo {author} {\bibfnamefont {A.}~\bibnamefont
  {Spracklen}}, \bibinfo {author} {\bibfnamefont {D.}~\bibnamefont
  {Choudhury}}, \bibinfo {author} {\bibfnamefont {N.}~\bibnamefont {Goldman}},
  \bibinfo {author} {\bibfnamefont {P.}~\bibnamefont {{\"O}hberg}}, \bibinfo
  {author} {\bibfnamefont {E.}~\bibnamefont {Andersson}}, \ and\ \bibinfo
  {author} {\bibfnamefont {R.~R.}\ \bibnamefont {Thomson}},\ }\bibfield
  {title} {\enquote {\bibinfo {title} {{Observation of a Localized Flat-Band
  State in a Photonic Lieb Lattice}},}\ }\href@noop {} {\bibfield  {journal}
  {\bibinfo  {journal} {Phys. Rev. Lett.}\ }\textbf {\bibinfo {volume} {114}},\
  \bibinfo {pages} {245504} (\bibinfo {year} {2015})}\BibitemShut {NoStop}%
\bibitem [{\citenamefont {Mikailov}\ \emph {et~al.}(2006)\citenamefont
  {Mikailov}, \citenamefont {{\c{S}}ent{\"u}rk}, \citenamefont {T{\"u}mbek},
  \citenamefont {Mammadov},\ and\ \citenamefont {Mammadov}}]{Mikailov:2006gg}%
  \BibitemOpen
  \bibfield  {author} {\bibinfo {author} {\bibfnamefont {F.~A.}\ \bibnamefont
  {Mikailov}}, \bibinfo {author} {\bibfnamefont {E.}~\bibnamefont
  {{\c{S}}ent{\"u}rk}}, \bibinfo {author} {\bibfnamefont {L.}~\bibnamefont
  {T{\"u}mbek}}, \bibinfo {author} {\bibfnamefont {T.~G.}\ \bibnamefont
  {Mammadov}}, \ and\ \bibinfo {author} {\bibfnamefont {T.~S.}\ \bibnamefont
  {Mammadov}},\ }\bibfield  {title} {\enquote {\bibinfo {title} {{Thermal
  hysteresis and memory effects in TlGaSe2 crystal with incommensurate
  phase}},}\ }\href@noop {} {\bibfield  {journal} {\bibinfo  {journal} {Phase
  Transitions}\ }\textbf {\bibinfo {volume} {78}},\ \bibinfo {pages} {413--419}
  (\bibinfo {year} {2006})}\BibitemShut {NoStop}%
\bibitem [{\citenamefont {Panich}(2008)}]{Panich:2008cs}%
  \BibitemOpen
  \bibfield  {author} {\bibinfo {author} {\bibfnamefont {A.~M.}\ \bibnamefont
  {Panich}},\ }\bibfield  {title} {\enquote {\bibinfo {title} {{Electronic
  properties and phase transitions in low-dimensional semiconductors}},}\
  }\href@noop {} {\bibfield  {journal} {\bibinfo  {journal} {J. Phys.: Condens.
  Matter}\ }\textbf {\bibinfo {volume} {20}},\ \bibinfo {pages} {293202}
  (\bibinfo {year} {2008})}\BibitemShut {NoStop}%
\end{thebibliography}
%

\end{document}